\documentclass[12pt]{article} 
\usepackage{amsmath,amssymb,tipx} 
\usepackage{graphicx} 
\usepackage{color} 
\usepackage{hyperref} 
\hypersetup{bookmarksopen=true, bookmarksnumbered=true, pdfstartview={FitH}, pdfborder={0 0 0}, colorlinks=true, linkcolor=red, linktoc=page, citecolor=blue, urlcolor=cyan, unicode=true}
\usepackage{geometry} 
\usepackage{fancyhdr} 
\usepackage[nottoc,notlof,notlot]{tocbibind} 
\usepackage[titles]{tocloft} 

\usepackage{parskip} 
\definecolor{title}{rgb}{0.1,0.5,0.9}
\definecolor{abst}{rgb}{0.366,0.366,0.266}
\definecolor{sect}{rgb}{1.0,0.0,0.0}
\definecolor{ssect}{rgb}{0.5,0.5,0.0}
\definecolor{sssect}{rgb}{0.33,0.33,0.33}
\definecolor{appsect}{rgb}{0.1,0.6,0.1}
\definecolor{ref}{rgb}{0.0,0.0,1.0}

\newcommand\sect[1] {{\color{sect}\section{#1}}}
\newcommand\subsect[1] {{\color{ssect}\subsection{#1}}}
\newcommand\subsubsect[1] {{\color{sssect}\subsubsection{#1}}}
\newcommand\appsect[1] {{\color{appsect}\section{#1}}}
\newcommand\references[1] {\color{ref} \begin{thebibliography}{[99]} \color{black} #1 \end{thebibliography}}
\numberwithin{equation}{section} 
\newcommand\bc {\begin{center}}
\newcommand\ec {\end{center}}
\newcommand\bnum {\begin{enumerate}}
\newcommand\enum {\end{enumerate}}
\newcommand\be {\begin{equation}}
\newcommand\ee {\end{equation}}
\newcommand\bfig {\begin{figure}}
\newcommand\efig {\end{figure}}
\newcommand\bpm {\begin{pmatrix}}
\newcommand\epm {\end{pmatrix}}
\renewcommand{\exp}{e^}
\renewcommand{\i}{\dot{\iota}}
\newcommand\A {{\cal A}}
\renewcommand\C {{\cal C}} 
\newcommand\D {{\cal D}}

\newcommand\M {{\cal M}}
\renewcommand\O {{\cal O}} 
\renewcommand\S {{\cal S}}
\renewcommand\' {^{\prime}}

\newcommand\y[2] {\frac{y_{#1}}{y_{#2}}}
\newcommand\cmp[3] {{\it Commun.\ Phys.\ Math.\ }{\bf #1} (#2) #3} 
\newcommand\cqg[3] {{\it Class.\ Quant.\ Grav.\ }{\bf #1} (#2) #3}
\newcommand\prl[3] {{\it Phys.\ Rev.\ Lett.\ }{\bf #1} (#2) #3}
\newcommand\pr[4] {{\it Phys.\ Rev.\ }{\bf #1 #2} (#3) #4} 
\newcommand\pl[4] {{\it Phys.\ Lett.\ }{\bf #1 #2} (#3) #4}
\newcommand\npb[3] {{\it Nucl.\ Phys.\ }{\bf B #1} (#2) #3}

\newcommand\arXivid[1] {\href{http://arxiv.org/abs/#1}{\tt arXiv:#1}}

\begin{document}
\pagenumbering{alph}
\title{\color{title}\Huge On Projective Hoops: Loops in Hyperspace}
\author{Dharmesh Jain\footnote{\href{mailto:djain@insti.physics.sunysb.edu}{djain@insti.physics.sunysb.edu}}\, , Warren Siegel\footnote{\href{mailto:siegel@insti.physics.sunysb.edu}{siegel@insti.physics.sunysb.edu}\vskip 0pt \hskip 8pt \href{http://insti.physics.sunysb.edu/\~siegel/plan.html}{http://insti.physics.sunysb.edu/$\sim$siegel/plan.html}}\bigskip\\ \emph{C. N. Yang Institute for Theoretical Physics}\\ \emph{State University of New York, Stony Brook, NY 11794-3840}}
\date{} 
\maketitle
\thispagestyle{fancy}
\rhead{YITP-SB-12-43} 
\lhead{\today}
\begin{abstract}
\normalsize We (re)derive the propagators and Feynman rules for the massless scalar and vector multiplets in N=2 Projective Superspace (`Projective Hyperspace'). With these, we are able to calculate both the divergent and finite parts of 2, 3 \& 4$-$point functions at 1-loop for N=2 Super-Yang-Mills theory (SYM) explicitly in Projective Hyperspace itself. We find that effectively only the coupling constant needs to be renormalized unlike in the N=1 case where an independent wavefunction renormalization is also required. This feature is similar to that of the background field gauge, even though we are using ordinary Fermi-Feynman gauge. The computation of 1-hoop beta-function is then straightforward and matches with the known result. We also show that it receives no 2-hoops contributions. All these calculations provide an alternative proof of the finiteness of N=4 SYM.
\end{abstract}

\newpage
\pagenumbering{Roman}
\cfoot{\thepage}\rhead{}\lhead{}
\tableofcontents

\newpage
\pagenumbering{arabic}
\sect{Introduction}
There has been a renewed interest in N=4 Super Yang-Mills theory (SYM) and its on-shell perturbative structure. These calculations mostly use components and / or some on-shell formulation of N=4 supersymmetry but often rely on unproven assumptions. It would be best to have an off-shell formalism for N=4 supersymmetry itself but it has been elusive for decades. The next best thing would be to use N=2 off-shell formalism which, as we show in this paper, is simpler than the well-known N=1 formalism.

Recently, we proposed the non-Abelian SYM action in N=2 Projective Superspace (Hyperspace) in \cite{DPHfH}. N=2 Supermultiplets (Hypermultiplets) in Projective Hyperspace have been long known since the work of Lindstr\"{o}m and Ro\v{c}ek\cite{ULMR}. The Feynman rules were derived for scalar and vector hypermultiplets in three successive papers by Gonzalez-Rey, et al\cite{FGR}. Some one-loop calculations involving scalar hypermultiplet's contributions to effective action were done in \cite{FGRc} but as the non-Abelian action was lacking, not much could be accomplished as far as calculations involving vector hypermultiplet were concerned.

Analogous (but slightly better) situation exists in the case of Harmonic Hyperspace developed by GIKOS\cite{GIKOS}. One-loop two-point functions for SYM effective action and four-point functions (both divergent \& finite) with external scalar hypermultiplets were computed by them in \cite{GIKOSc}. The $n-$point calculations were accomplished by Buchbinder, et al\cite{BB-ASef} but these are contributions to the effective action for the Abelian case only. Even a direct computation of the $\beta-$function for N=2 SYM has not been done, which requires a $3-$point calculation with ordinary Feynman rules. However, a 3$-$point calculation is unnecessary in the case of background field formalism, which does exist for Harmonic Hyperspace\cite{BB-Bg}. Using this formalism, even a 4$-$point S-matrix calculation in N=4 SYM has been done in \cite{BB-Bgc}, which also includes effective potential calculations similar to those in \cite{FGRc}.

In this paper, we extend the possible set of loop calculations in Projective Hyperspace and show that the hypergraphs are easier to handle than their N=1 counterparts! We calculate both the divergent and finite parts of 1-hoop 2, 3 \& 4$-$point functions. It turns out that the scalar hypermultiplet action (including its coupling to vector hypermultiplet) is not renormalized at any loop order. We also find that the divergent (and some finite) 1-loop corrections to SYM effective action have the same form as the classical action (modulo their momentum dependence) proving its renormalizability.

Both the wavefunction and coupling constant are linearly renormalized at 1-loop for N=2 SYM, which is not the case when N=1 supergraph methods are used\cite{PS, Storey}. An independent (non-linear) wavefunction renormalization is required in that case to keep the effective action renormalizable. Additionally, we learn from using hypergraph rules that there is effectively only one renormalization factor as is encountered when using background field formalism.\\

These 1-hoop calculations enable us to compute the well-known $\beta-$function for N=2 SYM coupled to scalar hypermultiplet (matter) in any representation of the gauge group. We also perform a few `selected' 2-hoops calculations to prove its two-loop finiteness. All these calculations and a few `miraculous' cancellations also show that the $\beta-$function of N=4 SYM vanishes at 1 \& 2-loop(s)\footnote{Using N=1 supergraph methods, finiteness of N=4 SYM has been shown till 3-loops explicitly in \cite{S3B0, CZ}. Using N=2 superfields and background field formalism, such cancellations leading to UV finiteness of N=2 \& 4 theories were explained in \cite{finite} for all loop orders.}. 

In the next section, we review the basics of Projective Hyperspace. After that, we write various hypermultiplet actions to derive the propagators and vertices, which enable us to present the revised `complete' Feynman rules to evaluate any possible hypergraph. Then, as mentioned above, we present some examples of 1 \& 2-hoop(s) hypergraph calculations and the resulting consequences for N=2 \& 4 theories.

\sect{Generic Theory}
We review the (relevant) generalities of Projective Hyperspace that are discussed in gory details in \cite{WS-AdS}.\\

\subsect{Hyperspace}
We start with $SU(2,2|2)$ group element ${g_{\M}}^{\A}$. The $SU(2)$ bosonic (Latin) and $SU(2,2)$ fermionic (Greek) indices contained in the group indices are divided into two parts and shuffled such that $\M=\{M,M\'\}=\{(m,\mu),(m\',\dot{\mu})\}$ with their values being $\{1,(1,2),1\',(\dot{1},\dot{2})\}$. Since the bosonic indices take only one value, they will be suppressed.

The projective coordinates $(4\,x's, 4\,\theta's\,\&\,1\,y)$ are arranged in an off-diagonal square matrix ${w_M}^{A\'}$ inside ${g_{\M}}^{\A}$. The rest of the fermionic coordinates $(\vartheta_{\mu},\vartheta^{\dot{\alpha}})$ are contained in the diagonal parts of $g$ and can be understood by the method of projection given below:
\begin{align}
g:&\quad{g_{\M}}^{\A}\rightarrow \bar{z}_{\M}{}^{A\'}\\
g^{-1}:&\quad{g_{\A}}^{\M}\rightarrow{z_{A}}^{\M}\\
Constraint:&\quad{z_A}^{\M}\bar{z}_{\M}{}^{A\'}=0\\
Solution:&\quad\left\{ \begin{array}{ll} \bar{z}_{\M}{}^{A\'}=\left({w_{M}}^{N\'},\delta_{M\'}^{N\'}\right)\bar{u}_{N\'}{}^{A\'}\,;\\ {z_A}^{\M}={u_A}^{N}\left(\delta_{N}^{M},-{w_N}^{M\'}\right). \end{array} \right.
\end{align}

The coordinates in $w,\,u,\,\&\,\bar{u}$ are arranged as follows: 
\begin{align}
{w_{M}}^{A\'}=&\bpm y & \bar{\theta}^{\dot{\alpha}} \\ \theta_{\mu} & {x_{\mu}}^{\dot{\alpha}}\epm\\
{u_{M}}^{A}=&\bpm I & 0 \\ \vartheta_{\mu} & I\epm\\
{\bar{u}_{M\'}}^{A\'}=&\bpm I & -\bar{\vartheta}^{\dot{\alpha}} \\ 0 & I\epm
\end{align}

These matrices have the following finite superconformal transformations (indices are suppressed in matrix notation below):
\begin{gather}
\bar{z}\'=g_0\bar{z},\quad z\'=z g_0^{-1};\quad g_0=\bpm a & b\\ c & d \epm,\quad g_0^{-1}=\bpm \tilde{d} & -\tilde{b}\\ -\tilde{c} & \tilde{a}\epm\\
\Rightarrow w\'=(aw+b)(cw+d)^{-1},\quad u\'=(w\tilde{c}+\tilde{d})^{-1}u,\quad \bar{u}\'=\bar{u}(cw+d)^{-1}\label{trwu}
\end{gather}

We can also construct symmetry invariants as differentials or finite differences:
\be
{z_A}^{\M}d\bar{z}_{\M}{}^{A\'}={u_A}^M\left({dw_M}^{M\'}\right)\bar{u}_{M\'}{}^{A\'},\quad {z_{2A}}^{\M}\bar{z}_{1\M}{}^{A\'}={u_{2A}}^M\left(w_1-w_2\right)_M{}^{M\'}\bar{u}_{1M\'}{}^{A\'}
\label{syminv}
\ee

\subsect{Covariant Derivatives}
It is easier to derive the symmetry generators $(G=g\partial_g)$ and covariant derivatives $(D=\partial_g g)$ from the infinitesimal forms of the transformations given above and in matrix form, they read:
\begin{gather}
G_w=\partial_w,\quad G_u=w\partial_w+u\partial_u,\quad G_{\bar{u}}=\partial_w w+\partial_{\bar{u}}\bar{u}\\
D_w=\bar{u}\partial_w u,\quad D_u=\partial_u u,\quad D_{\bar{u}}=\bar{u}\partial_{\bar{u}}
\end{gather}
 This defines the `projective representation', which is not quite useful for the construction of a `simple' N=2 SYM action. For that, we need what is called a `reflective representation' in which the $D$'s are `switched' with $G$'s. The explicit forms of covariant derivatives for all the coordinates in both representations are given in table \ref{cdrep}.
 \begin{table}[h]
\bc
\caption{Covariant Derivatives}
\vskip 2mm
\begin{tabular}{c|c|c}
$D$'s & Projective ($\check{\Pi}$) & Reflective ({\Large \textrevscr}) \\
\hline
$d_x$ & $\partial_x$ & $\partial_x$ \\
$d_{\theta}$ & $\partial_{\theta}-\bar{\vartheta}\partial_x$ & $\partial_{\theta}$ \\
$\bar{d}_{\theta}$ & $\partial_{\bar{\theta}}+\partial_x\vartheta$ & $\partial_{\bar{\theta}}$ \\
$d_y$ & $\partial_y-\bar{\vartheta}\partial_{\bar{\theta}}-\vartheta\partial_{\theta}-\bar{\vartheta}\partial_x\vartheta$ & $\partial_y$ \\
$d_{\vartheta}$ & $\partial_{\vartheta}$ & $\partial_{\vartheta}+y\partial_{\theta}+\bar{\theta}\partial_x$ \\
$\bar{d}_{\vartheta}$ & $\partial_{\bar{\vartheta}}$ & $\partial_{\bar{\vartheta}}+y\partial_{\bar{\theta}}+\partial_x\theta$\\
\end{tabular}
\label{cdrep}
\ec
\end{table}

All the $D-$commutators can be read directly from table \ref{cdrep} and are same in both the representations except the first one below, which is non-trivial only in {\Large \textrevscr}:
\begin{gather}
\{d_{1\vartheta},\bar{d}_{2\vartheta}\}=y_{12}d_x \\
\{d_{1\theta},\bar{d}_{2\vartheta}\}=d_x=\{\bar{d}_{1\theta},d_{2\vartheta}\} \\
[d_y,d_{\vartheta}]=d_{\theta}\quad\&\quad [d_y,\bar{d}_{\vartheta}]=\bar{d}_{\theta}
\end{gather}
The subscript `$a$' in $d_{a\vartheta}$'s labels different $y$'s (to condense notation, it will also label $\vartheta$'s wherever required!), $y_{12}\equiv y_1-y_2$ and $d_{\theta}\equiv d_{a\theta}$. All these commutations lead to the following useful identities\footnote{$d_{\theta}^4=d_{\theta}^2\bar{d}_{\theta}^2$, $d_{\theta}^2=\frac{1}{2}C_{\beta\alpha}d_{\theta}^\alpha d_{\theta}^\beta$, and so on.}:
\begin{gather}
d_{1\vartheta}d_{2\vartheta}^4=y_{12}d_{1\theta}d_{2\vartheta}^4 \\
d_{1\vartheta}^4d_{2\vartheta}^4=y_{12}^2\left[\frac{1}{2}\square+y_{12}d_{1\theta}d_x\bar{d}_{1\theta}+y_{12}^2d_{1\theta}^4\right]d_{2\vartheta}^4 \label{d4d4I}\\
\delta^8(\theta_{12})d_{1\vartheta}^4d_{2\vartheta}^4\delta^8(\theta_{21})=y_{12}^4\delta^8(\theta_{12})\\
d_{\vartheta}^4d_y^2d_{\vartheta}^4=\square d_{\vartheta}^4
\end{gather}

\subsect{Hyperfields}
We define a projective hyperfield $\Phi$ such that $d_{\vartheta}\Phi=\bar{d}_{\bar{\vartheta}}\Phi=0$. In $\check{\Pi}$, it just means that $\Phi\equiv \Phi(x,\theta,\bar{\theta},y)$. This representation is useful for defining actions in projective hyperspace. In {\Large \textrevscr}, the dependence on $(\vartheta\,\&\,\bar{\vartheta})$ is non-trivial and looks like: $\Phi\equiv \Phi(x+\vartheta\bar{\theta}+\theta\bar{\vartheta}-y\vartheta\bar{\vartheta},\theta-y\vartheta,\bar{\theta}-y\bar{\vartheta},y)$. This representation is more suited for writing actions in the `full' hyperspace with $8\,\theta$'s.

The superconformal transformation of $\Phi$ with a (superscale) weight `$\omega$' can be deduced by requiring that $dw\,\Phi^{1/\omega}$ transforms as a scalar. The resulting transformations are:
\be
dw\'=dw[sdet(cw+d)]^{2},\quad \Phi(w\')=[sdet(cw+d)]^{-2\omega}\Phi(w)
\ee 
This means that the Lagrangian should have $\omega=1$ for the action to be superconformally invariant. An example of this will be the scalar hypermultiplet action.

Charge conjugate expressions of the coordinates can be derived in a way similar to the derivation of superconformal transformations:
\begin{align}
\C\bpm y & \bar{\theta} \\ \theta & x\epm^{\dagger}=&\bpm -\frac{1}{y} & \frac{\bar{\theta}}{y} \\ \frac{\theta}{y} & x-\frac{\theta\bar{\theta}}{y}\epm\\
\C(\vartheta)^{\dagger}=&\,\vartheta-\frac{\theta}{y}\\
\C(\bar{\vartheta})^{\dagger}=&\,\bar{\vartheta}+\frac{\bar{\theta}}{y}
\end{align}
The conjugate of the hyperfield $\Phi$ can be now defined as follows:
\be
\C(\Phi)^{\dagger}=y^{2\omega}[\Phi(\C w)]^{\dagger}
\ee

\subsect{Internal Coordinate}
Much of projective hyperspace can be understood by analogy to full N=1 superspace, as a consequence of both having 2 $\theta$'s and 2 $\bar{\theta}$'s. Then what we've left to understand is the treatment of the internal $y-$coordinate. The field strengths turn out to be Taylor expandable in $y$ on-shell\cite{WS-AdS}, so their charge conjugates must be Laurent expandable on-shell. Thus, it seems natural to use contour integration:
\be
\oint \frac{dy}{2\pi\i}\frac{y^m}{y^n}=\delta_{m+1,n}
\ee
(The factor of $2\pi\i$ will be suppressed in what follows.) This makes the $y-$space effectively compact, as expected for an internal symmetry. It is also a convenient way to constrain a generic hyperfield $\Phi$'s $y-$dependence:
\be
\phi(y)\left[0^{\uparrow}\right] = \oint_{0,y}dy\'\frac{1}{y\'-y}\Phi(y')\left[0_{\downarrow}^{\uparrow}\right] = \sum_{n=0}^{\infty}y^n\oint_{0}dy\'\frac{1}{{y\'}^{n+1}}\Phi(y')\left[0_{\downarrow}^{\uparrow}\right]
\ee
Here, $\phi(y)$ has only the non-negative powers of $y$ encoded in the notation $\left[0^{\uparrow}\right]$. The coefficients of different powers of $y$ in $\phi$ matches with the correct ones in $\Phi(y\')$ which has all the powers of $y$ denoted by $\left[0_{\downarrow}^{\uparrow}\right]$. Thus, the contour integral acts as an `arctic' projector and $\phi$ is an arctic hyperfield, being regular at origin.

As for Feynman diagrams in Minkowski space, it is often more convenient, when defining how to integrate around poles (especially when there's more than one integral to evaluate), to move the poles rather than the contour. In this interpretation, instead of having a bunch of integrals over various contours, with the poles for integration over each variable lying on the contour of another variable, we have all integrals over the same contour, with all poles in various different positions near that contour. For our case, the appropriate `$\epsilon-$prescription' is given by writing the arctic projection of $\Phi$ as
\be
\phi(y_2)\left[0^{\uparrow}\right] =\int dy_1\frac{1}{y_{12}}\Phi(y_1)\left[0_{\downarrow}^{\uparrow}\right],\quad \frac{1}{y_{12}}\equiv\frac{1}{y_1-y_2 + \epsilon(y_1 + y_2)}
\ee
at least for the case of any convex contour (e.g., a circular one) about the origin; otherwise, we need to invent a fancier notation. Similarly, an `antarctic' projector with the same $\epsilon-$prescription can be written for an antarctic hyperfield,
\be
\bar{\phi}(y_2)\left[(-1)_{\downarrow}\right]=\int dy_1\frac{1}{y_{21}}\Phi(y_1)\left[0_{\downarrow}^{\uparrow}\right]
\ee
where, $\left[(-1)_{\downarrow}\right]$ denotes $\bar{\phi}$ contains all the negative powers of $y$.

All these generalities now enable us to properly see them in action!\\

\sect{Specific Theory}
We start with writing the actions for various hypermultiplets and end with enumerating the Feynman rules, which allow us to do all the necessary calculations presented in the next section.\\

\subsect{Actions}
\paragraph{Scalar Hypermultiplet\\}
For the scalar hypermultiplet, the requirement of Laurent expandability in $y$ turns out to be too weak off-shell; we therefore require that it be Taylor expandable. This `polarity' (i.e. arctic or antarctic) will be the analog of the `chirality' of N=1 supersymmetry. Unlike the N=1 case, we now have an infinite number of auxiliary component fields because of the infinite Taylor expansion in $y$. The free action can be written in analogy to N=1 as:
\be
{\S}_{\Upsilon}=-\int dx\,d^4\theta\,dy\, \bar{\Upsilon}\Upsilon.
\label{sY}
\ee

For superconformal invariance and reality, the arctic hyperfield $\Upsilon\left[0^{\uparrow}\right]$ must have $\omega=\frac{1}{2}$. Its conjugate is an (almost) antarctic hyperfield $\bar{\Upsilon}\left[1_{\downarrow}\right]=y[\Upsilon(\C w)]^{\dagger}$. Note that the integral of $\Upsilon^2$ or $\bar{\Upsilon}^2$ would give $0$, just as for N=1, but now because of polarity rather than chirality. Also, since there is no analog to the chiral superpotential terms of N=1, there are no renormalizable self-interactions for this hypermultiplet. All its interactions will be through coupling to the vector hypermultiplet.

There is not much to say about the off-shell components: they are just the coefficients of Taylor expansion in $y$ and the $\theta$'s. So we examine the field equations to see how only a finite number of components survive on-shell. A direct and easy way to accomplish that is to use reflective representation. Using the 4 extra $\vartheta$'s, we can write both the arctic \& antarctic hyperfields in terms of an unconstrained (in both $y$ and $\vartheta$) hyperfield:
\begin{gather}
\Upsilon(y_2)\left[0^{\uparrow}\right] = d_{2\vartheta}^4\int dy_1\frac{1}{y_{12}}\Phi(y_1)\left[0_{\downarrow}^{\uparrow}\right] \label{y}\\
\bar{\Upsilon}(y_2)\left[1_{\downarrow}\right] = d_{2\vartheta}^4 d_{y_2}^2\int dy_1\frac{1}{y_{21}}\bar{\Phi}(y_1)\left[0_{\downarrow}^{\uparrow}\right] \label{Cy}
\end{gather}

The $y-$derivatives appear in \ref{Cy} because: (1) the antarctic projection makes `$-1$' the highest power of $y$; (2) the $y$ term in each $d_{2\vartheta}$ increases this to `$3$'; and (3) the two $y-$derivatives decrease this to the correct power of `$1$'. Unconstrained variation of the action with respect to $\bar{\Phi}$ (after using the $d_{\vartheta}^4$ to turn $\int d^4 \theta$ into $d^8 \theta$) then gives the field equations $d_y^2 \Upsilon=0$ (the arctic projection is redundant). On the other hand, variation with respect to $\Phi$ kills the antarctic pieces of $\bar{\Upsilon}$, which is the same as $d_y^2\bar{\Upsilon} = 0$. Due to superconformal invariance, the rest of the superconformal equations are also satisfied.

Thus, the on-shell component expansion of the scalar hypermultiplet reads\footnote{The $\theta\bar{\theta}-$term in $\Upsilon$ can be understood as a consequence of one of the superconformal field equations\cite{WS-AdS}, which schematically reads: $\partial_x\partial_y+\partial_{\theta}\partial_{\bar{\theta}}=0$.}:
\begin{align}
\Upsilon(x,\theta,\bar{\theta},y)=&(A+yB)+\theta\chi+\bar{\theta}\bar{\tilde{\chi}}-\theta\partial B\bar{\theta}\\
\bar{\Upsilon}(x,\theta,\bar{\theta},y)=&y\Biggl[\bar{A}-\frac{\theta \partial\bar{A}\bar{\theta}}{y}+\frac{\theta^2\bar{\theta}^2\square\bar{A}}{y^2}-\frac{\bar{B}}{y}+\frac{\theta^2\bar{\theta}^2\square\bar{B}}{y^3}+\frac{\theta}{y}\left(\chi-\frac{\theta \partial\chi\bar{\theta}}{y}\right)+\frac{\bar{\theta}}{y}\left(\bar{\tilde{\chi}}-\frac{\theta \partial\bar{\tilde{\chi}}\bar{\theta}}{y}\right)\Biggr]
\end{align}
From the last equation, we clearly see that the equations of motion for the complex scalars and the Weyl spinors are satisfied if $d_y^2\bar{\Upsilon} = 0$ applies.

\paragraph{Vector Hypermultiplet\\}
Like the scalar hypermultiplet, we look for a description of the vector hypermultiplet in terms of a prepotential defined on projective hyperspace. Again in analogy to N=1, this should be a real prepotential, rather than a polar one. Because it lacks the polarity restriction, and is thus Laurent expandable in $y$, it is called `tropical'. Like the scalar hypermultiplet, it will have only a few powers of $y$ surviving on-shell.

Just as for both N=0 \& 1, gauge symmetry is understood as a generalization of global symmetry, so we derive its form by coupling to matter. The straightforward generalization of the N=1 coupling is then given by the action for the scalar hypermultiplet coupled to a vector hypermultiplet background:
\be
{\S}_{\Upsilon-V}=-\int dx\,d^4\theta\,dy\, \bar{\Upsilon}\exp{V}\Upsilon.
\label{sYV}
\ee

This coupling fixes the weight of $V$ to be $0$:
\be
V\'(w) = V(w\'),\quad \bar{V}(w)\equiv [V(\C w)]^{\dagger} =V(w)
\ee
The gauge transformations are then
\be
\Upsilon\'=\exp{\i\Lambda}\Upsilon,\quad \bar{\Upsilon}\'=\bar{\Upsilon}\exp{-\i\bar{\Lambda}},\quad {\exp{V}}\'=\exp{i\bar{\Lambda}}\exp{V}\exp{-\i \Lambda}
\ee
where $\Lambda$ is arctic like $\Upsilon$, but has $\omega=0$ like $V$. Thus, $\bar{\Lambda}$ has only non-positive powers of $y$, unlike $\bar{\Upsilon}$. Because of the $1/y$'s associated with conjugated coordinates, setting $\Lambda$ to $\bar{\Lambda}$ would reduce $\Lambda$ to a real constant, i.e. the global symmetry.

With this gauge invariance, we can examine the on-shell component fields of the vector hypermultiplet. Since $\Lambda$ contains all non-negative powers of $y$, and $\bar{\Lambda}$ contains all non-positive powers, it might seem that everything can be gauged away, but again the additional $1/y$'s associated with charge conjugation modify things: The $1/y$ in $\C \theta$ increases the number of non-gauge components of V with increasing $\theta$, while the $\theta\bar{\theta}/y$ in $\C x$ leads to an $x-$derivative gauge transformation, again in analogy with the N=1 case. (We can also look at just what $\Lambda$ gauges away, and then apply `reality' to $V$.) The result is that, unlike the scalar hypermultiplet (but like the N=1 vector multiplet), $V$ has a finite number of auxiliary fields.

In a Wess-Zumino gauge,
\be
V = \frac{1}{y}\left[\left(\theta A\bar{\theta}+\theta^2\phi+\bar{\theta}^2\bar{\phi}\right)+\bar{\theta}^2\theta\left(\lambda+\frac{\tilde{\lambda}}{y}\right)+\theta^2\bar{\theta}\left(\bar{\lambda}+\frac{\bar{\tilde{\lambda}}}{y}\right)+\theta^2\bar{\theta}^2\left(\D+\frac{\D_0}{y}+\frac{\bar{\D}}{y^2}\right)\right]
\ee
where the residual gauge invariance is the usual one for the vector $A$. We thus find, in addition to the expected physical vector ($A$), a complex scalar ($\phi$) and SU(2) doublet of spinors ($\lambda\,\&\,\tilde{\lambda}$), there is an SU(2) triplet of auxiliary scalars ($\D,\bar{\D}\,\&\,\D_0$). This same set of fields is found if the vector hypermultiplet is reduced to N=1 supermultiplets, one vector supermultiplet plus one scalar supermultiplet. In the N=1 case, the construction of the vector multiplet action depended on the fact that a spinor derivative could kill the chiral gauge parameter. In the N=2 case, we have arctic and antarctic gauge parameters, and the only way to kill them is by antarctic or arctic projection. This leads to an action of the form
\be
{\S}_{V}=\frac{tr}{g^2}\int dx\,d^8\theta\sum_{n=2}^{\infty}\frac{(-1)^n}{n}\prod_{i=1}^{n}\int dy_i\frac{\left(\exp{V_1}-1\right)\left(\exp{V_n}-1\right)}{y_{12}y_{23}...y_{n1}}
\label{sV}
\ee
where, $V_i\equiv V(x,\theta,\vartheta,y_i)$. This action is invariant under the following gauge transformation (details are in \cite{DPHfH}):
\be
\delta\left(\exp V\right)=\i\left(\exp V \Lambda -\bar{\Lambda}\exp V\right)\Rightarrow \delta V=\i\left[\frac{V}{2},\left((\Lambda+\bar{\Lambda})+\left[\mathrm{coth}\frac{V}{2},(\Lambda-\bar{\Lambda})\right]\right)\right]. 
\ee

Superconformal invariance of the action might not be obvious, especially because of the non-locality. The first thing to note is that the full superspace volume element ($\int dx\,d^8\theta$) is superconformally invariant (because $sdet(g_0)=1$). Next is to use the results for the superconformal transformations of $dw_i$ and $w_{ij}$, read from \ref{trwu} \& \ref{syminv}, to find those for the coordinate $y$:
\begin{align}
dy\'_i=&\,\frac{dy_i}{(w_i\tilde{c}+\tilde{d})(cw_i+d)}\\
y\'_{ij}=&\,\frac{y_{ij}}{(w_i\tilde{c}+\tilde{d})(cw_j+d)}
\end{align}
where, the factors $(cw_i+d)$, etc denote the single matrix element corresponding to the $y-$coordinate. We also use the fact that the other $w_{ij}$'s vanish as the action is local in these coordinates. The similar transformation factors of $dy_i$'s \& $y_{ij}$'s then cancel due to the `cyclic' nature of the denominator in SYM action proving its superconformal invariance.

\paragraph{Ghost Hypermultiplets\\}
The introduction of ghosts follows the usual BRST procedure and is analogous to the case of N=1 at least in the Fermi-Feynman gauge (see section \ref{P} for some details.):
\begin{align}
{\S}_{bc}=&-tr\int dx\,d^4\theta\,dy\,(y\,b+\bar{b})\left[\frac{V}{2},\left(\left(c+\frac{\bar{c}}{y}\right)+\left[\mathrm{coth}\frac{V}{2},\left(c-\frac{\bar{c}}{y}\right)\right]\right)\right]\nonumber\\
=&-tr\int dx\,d^4\theta\,dy\,\left[\bar{b}\,c+\bar{c}\,b+(y\,b+\bar{b})\frac{V}{2}\left(c+\frac{\bar{c}}{y}\right)+\frac{1}{3}(y\,b+\bar{b})\frac{V^2}{4}\left(c-\frac{\bar{c}}{y}\right)+...\right]
\label{sghf}
\end{align}

We can also choose a non-linear gauge like the Gervais-Neveu gauge in which the ghost action would be simplified to:
\begin{align}
{\S}_{bc}=&-tr\int dx\,d^4\theta\,dy\,(y\,b+\bar{b})\left[\exp V c-\frac{\bar{c}}{y}\,\exp V\right]\nonumber\\
=&-tr\int dx\,d^4\theta\,dy\,\left[y\,b\,\exp V c+\bar{c}\,\exp V b+\bar{b}\,\exp V c+\frac{1}{y}\bar{c}\,\exp V\bar{b}\right]
\label{sghg}
\end{align}

There does not seem to be any real advantage of this gauge (e.g. to show the non-renormalization of $g$ in N=4 SYM is not that straightforward) apart from the absence of `weird' numerical factors coming from the expansion of $\mathrm{coth}(x)$ in the case of Fermi-Feynman gauge. So we will use action \ref{sghf} in all the calculations presented later.

\subsect{Propagators\label{P}}
\paragraph{Scalar\\}
We add source terms to the quadratic action of $\Upsilon$ and convert the $d^4\theta$ integral to $d^8\theta$ integral by rewriting $\Upsilon$'s using equations \ref{y} \& \ref{Cy}\footnote{Writing $\Upsilon$ instead of $\Phi$ does not make a difference here.}:
\begin{align}
{\S}_{\Upsilon-J}=&-\int dx\, d^4\theta\, dy \left[\bar{\Upsilon}\Upsilon+\bar{J}\Upsilon+\bar{\Upsilon}J\right]\label{actU}\\
=&-\int dx\, d^8\theta\int dy_1\left[d_{y_1}^2\int dy_3\frac{\bar{\Upsilon}_3}{y_{13}}d_{1\vartheta}^4\int dy_2\frac{\Upsilon_2}{y_{21}}+\bar{J}_1\int dy_2\frac{\Upsilon_2}{y_{21}}+d_{y_1}^2\int dy_3\frac{\bar{\Upsilon}_3}{y_{13}}J_1\right]\nonumber
\end{align}

The sources $J\,\&\,\bar{J}$ are generic projective hyperfields with $\omega=\frac{1}{2}$. Now, the modified equations of motion of $\bar{\Upsilon}\,\&\,\Upsilon$ can be derived from above and (after some integration by parts) they read:
\begin{align}
\int dy_1 \frac{d_{1\vartheta}^4 d_{y_1}^2\Upsilon_1}{y_{13}}=&-\int dy_1 d_{1\vartheta}^4 d_{y_1}^2\left(\frac{1}{y_{13}}\right)J_1\nonumber\\
\Rightarrow \square \Upsilon_3=&-d_{3\vartheta}^4\int dy_1 \frac{2J_1}{y_{13}^3} \label{eomU}\\
\mathrm{Similarly,}\,\,\square\bar{\Upsilon}_2=&-d_{2\vartheta}^4\int dy_1 \frac{2\bar{J}_1}{y_{21}^3}\label{eomUb}
\end{align}
Plugging the equations \ref{eomU} \& \ref{eomUb} back in action \ref{actU}, we get:
\begin{align}
{\S}_{\Upsilon-J}=&\,\frac{1}{2}\int dx\,d^8\theta\,dy_1\left[\bar{J}_1\frac{1}{\frac{1}{2}\square}\int dy_2 \frac{J_2}{y_{21}^3}+\frac{1}{\frac{1}{2}\square}\int dy_2 \frac{\bar{J}_2}{y_{12}^3}J_1\right]\nonumber \\
=&\int dx\,d^8\theta\,dy_1\,dy_2\left[\bar{J}_1\frac{1}{y_{21}^3}\frac{1}{\frac{1}{2}\square}J_2\right]
\label{SYJ}
\end{align}
This gives us the following scalar propagator:
\be
\langle\Upsilon(1)\bar{\Upsilon}(2)\rangle=-\frac{d_{1\vartheta}^4d_{2\vartheta}^4\delta^8(\theta_{12})}{y_{21}^3}\frac{\delta(x_{12})}{\frac{1}{2}\square}.
\ee

\paragraph{Vector\\}
Gauge fixing of the vector hypermultiplet action looks similar to the N=1 case, in the same sense that the scalar hypermultiplet action does. The main modifications are that now $d^4\theta$ is projective, there is also $dy$, the ghosts and Nakanishi-Lautrup fields are projective arctic / antarctic fields instead of chiral / anti-chiral ones. The $y-$dependence of ghosts $c\,\&\,\bar{c}$ is $\left[0^\uparrow\right]\,\&\,\left[0_\downarrow\right]$; anti-ghosts $b\,\&\,\bar{b}$ is $\left[0^\uparrow\right]\,\&\,\left[2_\downarrow\right]$ and NL fields $B\,\&\,\bar{B}$ is $\left[0^\uparrow\right]\,\&\,\left[2_\downarrow\right]$. We redefine the conjugates so that their $y-$dependence is similar to $\bar{\Upsilon}$:
\be
\bar{c}\left[0_\downarrow\right]\rightarrow\frac{1}{y}\bar{c}\left[1_\downarrow\right];\, \bar{b}\left[2_\downarrow\right]\rightarrow y\,\bar{b}\left[1_\downarrow\right];\, \bar{B}\left[2_\downarrow\right]\rightarrow y\,\bar{B}\left[1_\downarrow\right]
\ee

We choose the following gauge-fixing function:
\begin{align}
V_{gf}=&\int dx\,d^4\theta\,dy\,(y\,b+\bar{b})V\\
\delta V_{gf}=&\int dx\,d^4\theta\,dy \left[(y\,B+\bar{B})V+(y\,b+\bar{b})\delta V \left(c,\frac{\bar{c}}{y}\right)\right]
\end{align}
The second term gives ${\cal S}_{bc}$ in Fermi-Feynman gauge\footnote{Choosing $\left(\exp V-1\right)$ instead of $V$ in $V_{gf}$ gives the ghost action \ref{sghg}.} (eq. \ref{sghf}). The first term along with a gauge-averaging term (kinetic term for NL field) gives us the gauge-fixing action:
\begin{align}
{\S}_{gf}=&\,\frac{tr}{g^2}\int dx\,d^4\theta\,dy\left[-\bar{B}\frac{1}{\square}B+(y\,B+\bar{B})V\right]\\
\Rightarrow{\cal S}_{gf}=&\,\frac{tr}{2g^2}\int dx\,d^8\theta\,dy_1\,dy_2\,V_1\left[\frac{y_2}{{y_{12}}^3}+\frac{y_1}{{y_{21}}^3}\right]V_2
\end{align}
The final expression for $\S_{gf}$ follows from similar manipulations employed in deriving eq. \ref{SYJ}, i.e. by integrating out $B$ \& $\bar{B}$ using their equations of motion.

We now combine the terms quadratic in $V$ from the above equation and eq. \ref{sV} to get:
\begin{align}
{\S}_{V}^{(2)}+{\cal S}_{gf}^{(2)}=&-\frac{tr}{2g^2}\int dx\,d^4\theta\,dy_1\,dy_2\,V_1\frac{1}{y_{12}^2}\left[1-\frac{y_2}{y_{12}}-\frac{y_1}{y_{21}}\right]d_{1\vartheta}^4 V_2\nonumber\\
=&\,\frac{tr}{2g^2}\int dx\,d^4\theta\,dy_1\,dy_2\,V_1\frac{1}{y_{12}^2}\left[\frac{y_1+y_2}{2}\delta(y_{12})\right]y_{12}^2\left(\frac{1}{2}\square+\O(y_{12})\right) V_2\nonumber\\
=&\,\frac{tr}{2g^2}\int dx\,d^4\theta\,dy\,y\,\frac{V\square V}{2}
\end{align}
This gives the following vector propagator:
\be
\langle V(1)V(2)\rangle=d_{1\vartheta}^4\delta^8(\theta_{12})\frac{\delta(y_{12})}{y_1}\frac{\delta(x_{12})}{\frac{1}{2}\square}.
\ee

\paragraph{Ghosts\\}
The derivation of ghost propagators proceeds along similar lines to that of the scalar propagator and the results are:
\begin{align}
\langle\bar{b}(1)c(2)\rangle=\langle\bar{c}(1)b(2)\rangle=&\,\frac{d_{2\vartheta}^4d_{1\vartheta}^4\delta^8(\theta_{12})}{y_{12}^3}\frac{\delta(x_{12})}{\frac{1}{2}\square}\,,\\
\langle c(1)\bar{b}(2)\rangle=\langle b(1)\bar{c}(2)\rangle=&-\frac{d_{1\vartheta}^4d_{2\vartheta}^4\delta^8(\theta_{12})}{y_{21}^3}\frac{\delta(x_{12})}{\frac{1}{2}\square}.
\end{align}

\subsect{Vertices\label{V}}
\paragraph{$\Upsilon$\\}
The scalar hypermultiplet does not have any self-interactions. Only $\Upsilon-V$ vertices are possible as is evident from the actions written above (We use the group theoretical conventions and diagrams along the lines of \cite{S3B0}\footnote{To summarize: The vector \& ghost hyperfields are in the adjoint representation of gauge group and the scalar hyperfield is in some representation $R$: $V=V^aT_a,\,\Upsilon=\Upsilon^aT_a,$ etc. The group generators ($T_a$) satisfy $[T_a,T_b]=\i{f_{ab}}^cT_c$ and in adjoint rep: ${(T_a)_b}^c=\i {f_{ab}}^c=\,\line(1,0){10}\line(0,1){12}\line(-1,0){4}\line(1,0){17}\,$. The Dynkin index ($c_A$) is defined by: $tr_A(T_aT_b)=f_{acd}{f_b}^{cd}=c_A\delta_{ab}$. In $R$, this trace generalizes to: $tr_R(T_aT_b)=c_R\delta_{ab}$.}.):
\[\bar{\Upsilon}^i V^{j_1}...V^{j_n}\Upsilon^k \,\rightarrow\,\int d^4\theta\int dy\left({}_i\line(1,0){15}\line(0,1){15}^{j_1}\line(-1,0){8}\line(1,0){10}...\line(1,0){10}\line(0,1){15}^{j_n}\line(-1,0){9}\line(1,0){15}{}_k\right)\]
where, the group theory factor (shown in parentheses) is for adjoint representation.

\paragraph{$V$\\}
Pure vector hypermultiplet vertices take the following form:
\[(V_1)^{m_1}...(V_n)^{m_n} \,\rightarrow\,\int d^8\theta\,dy_1\,...\,dy_n \frac{1}{y_{12}...y_{n1}}\left({}_1\line(1,0){15}\line(0,1){15}\line(1,0){8}...\line(1,0){8}\line(0,1){15}\line(1,0){15}{}_n\right)\]
The group theory factor shown above corresponds to the case of $m_1=...=m_n=1$. For other cases, this factor depends on the number of $V$'s rather than that of the independent $y-$coordinates. Apart from this subtlety, the factor is still similar to the simplest case but we will not consider diagrams containing such vertices (with $m_i>1$) here.

\paragraph{$(b,c)$\\}
There are altogether four possibilities for ghost vertices and they differ in the accompanying $y-$integrals:
\begin{align*}
b\,V^n\,c\,&\rightarrow\,\int d^4\theta\int dy\,y\\
\bar{c}\,V^n\,b\,&\rightarrow\,\int d^4\theta\int dy\\
\bar{b}\,V^n\,c\,&\rightarrow\,\int d^4\theta\int dy\\
\bar{c}\,V^n\,\bar{b}\,&\rightarrow\,\int d^4\theta\int dy\,\frac{1}{y}
\end{align*}
Group theory factors for these ghost vertices are similar to those for the scalar vertices.

\subsect{Feynman Rules}
\bnum
\item Basic set-up: Apply usual Feynman rules for drawing diagrams and writing expressions for them using the propagators \& vertices derived above.

\item $d-$Algebra: Convert $d^4\theta$ integrals to $d^8\theta$ integrals by taking $d_{a\vartheta}^4$'s off the propagators. There should be at least two $d_{a\vartheta}^4$'s remaining for the diagram to not vanish. Remove the remaining $d_{a\vartheta}^4$'s using integration by parts (which implicitly assumes using the `freed' $\delta^8(\theta_{12})$ to do one $d^8\theta$ integral) and keep using the identity \ref{d4d4I} till only one $d_{a\vartheta}^4$ is left\footnote{All this can be summarized by the formula: $n_{\theta}-\left(n_{\delta}-\frac{n_{\vartheta}}{2}\right)=\frac{1+n}{2}$; where, $n_{\theta}$=no. of $\int d^8\theta$, $n_{\delta}$=no. of $\delta^8(\theta)$, $n_{\vartheta}$=no. of $d^4_{a\vartheta}$ and $n$=no. of times the identity \ref{d4d4I} has to be applied, which means that a diagram vanishes if $n\leq 0$.}. As far as computing divergences is concerned, this leads to a deceptively simple result for a 1-hoop diagram (or a particular 1-hoop in a multi-hoop diagram):
\be
d_{a\vartheta}^4d_{b\vartheta}^4...d_{z\vartheta}^4=\frac{1}{2}k^2 y_{ab}^2 y_{b\cdot}^2...y_{\cdot z}^2 y_{za}^2 d_{z\theta}^4 d_{z\vartheta}^4
\ee 
where, $k$ is the loop momentum and the second-to-last $\theta-$integral can be done by using this identity: $\delta^8(\theta_{12})d_\theta^4d_{\vartheta}^4\delta^8(\theta_{12})=\delta^8(\theta_{12})$.

\item $y-$Calculus: Use the identities in Appendix \ref{App:yC} to do `some' of the $y-$integrals. Specifically, for evaluation of divergences, perform partial fractions to generate the cyclic $y-$denominator $(y_{12}y_{23}...y_{n1})$ of the SYM action. Then, performing the extra $y-$integrals is equivalent to just replacing the extra $y$'s in the integrand by following these two rules: (a) Remove the factor $\int \frac{dy_a}{y_{a1}}$ after replacing all non-negative powers of $y_a$ by $y_1$ and setting its negative powers to $0$; (b) Remove the factor $\int \frac{dy_a}{y_{1a}}$ after replacing all negative powers of $y_a$ by $y_1$ and setting other (non)-occurrences of $y_a$ to $0$.

\item Miscellaneous: Evaluate group theoretical factors and track down signs \& symmetry factors. Finally, evaluate the integrals over loop-momenta.
\enum

\sect{Results}

\subsect{1-hoop Examples}

\subsubsect{Scalar Self-energy}
There are two diagrams with different propagators making the loop as shown in figure \ref{S1H2P}.
\bfig[h]\bc
\includegraphics[width=8cm,height=2cm]{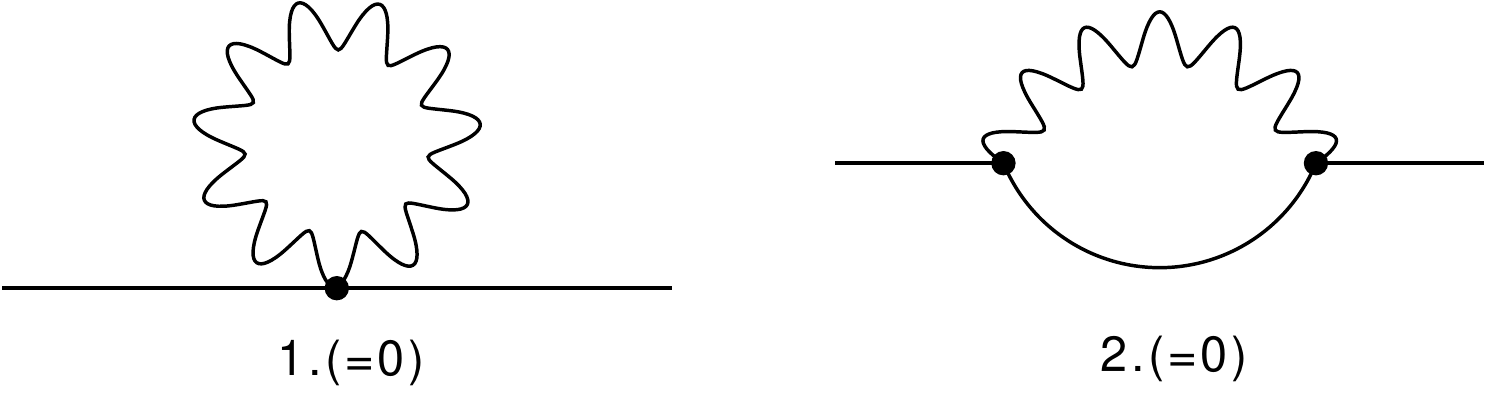}
\caption{Scalar self-energy diagrams at 1-hoop.}
\label{S1H2P}
\ec\efig
\bnum
\item $1\, V-$propagator: Remove $d^4_{\vartheta}$ from the vector propagator to get the $d^8\theta$ measure. This leaves no $d_\vartheta$\rq{}s to kill the $\delta^8(\theta_{12})$, so this diagram vanishes. Such tadpole diagrams (even multi-hoop diagrams containing these as sub-diagrams) always vanish, so we will not consider these anymore in what follows.

\item $1\,V-\,\&\,1\,\Upsilon-$propagators: Remove two $d^4_{\vartheta}$\rq{}s from these propagators to complete the two $d^4\theta$ measures. This means there are not enough (in fact, only $4$) $d_\vartheta$'s left to kill one of the $\delta^8(\theta_{12})$, so this diagram also vanishes.
\enum

This means that the scalar hyperfield is not renormalized which is obvious from the fact that its action is over only the projective hyperspace but the Feynman diagrams give contributions over full hyperspace. In other words, scalar hypermultiplet action (coupled to vector hypermultiplet, as shown below) is not renormalized at any loop order.

\subsubsect{\texorpdfstring{$\bar{\Upsilon}\,V\,\Upsilon$}{Scalar-Vector 3-point Function}}
There are four diagrams in this case as shown in figure \ref{YVY1H}. Two of these diagrams vanish because of $d-$algebra similar to the self-energy case. The other two are evaluated below:
\bfig[h]\bc
\includegraphics[width=14cm,height=2.5cm]{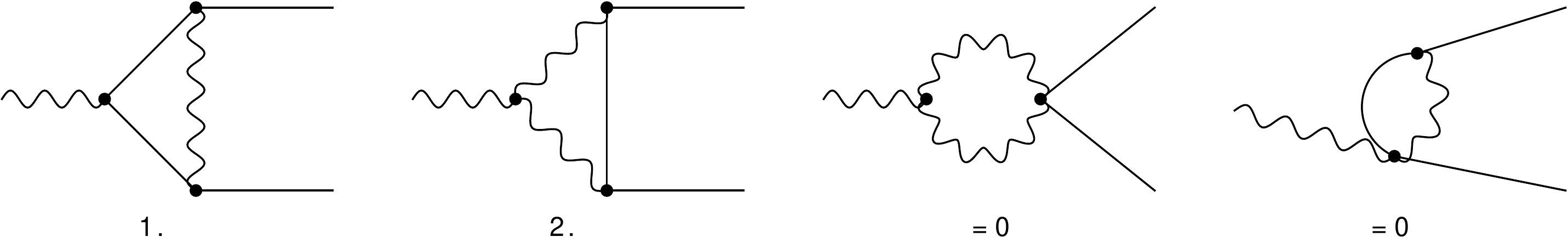}
\caption{$\bar{\Upsilon}\,V\,\Upsilon$ diagrams at 1-hoop.}
\label{YVY1H}
\ec\efig
\bnum
\item $1\,V-\,\&\,2\,\Upsilon-$propagators: This diagram has enough (8, at last!) $d_\theta$'s to kill one of the $\delta(\theta)$ functions so that two $\theta-$integrals can now be done. However, this generates only a $y_{12}^4-$factor without any momentum factors in the numerator, which makes this diagram power-counting finite and the explicit finite result reads:
\be
-\frac{c_A}{2}\int\frac{d^4k}{(2\pi)^4} \frac{1}{\frac{1}{2}k^2\frac{1}{2}(k+p_2)^2\frac{1}{2}(k-p_1)^2}\int d^8\theta\int \frac{dy_{1,2}}{y_2}\frac{\bar{\Upsilon}_2(p_2) V_1(p_1){\Upsilon}_2(p_3)}{y_{12}\,y_{21}}\label{A3h}
\ee
where, $p_i$'s are the external momenta.

\item $2\,V-\,\&\,1\,\Upsilon-$propagators: Applying similar maneuvers as above, we conclude that this diagram is also finite:
\be
-\A_3(p_2,-p_1)\times\frac{c_A}{2}\int d^8\theta\int \frac{dy_{1,2,3}}{y_2\,y_3}\frac{\bar{\Upsilon}_2(p_2)V_1(p_1){\Upsilon}_3(p_3)}{y_{12}\,y_{31}}
\ee
where, $\A_3(p_2,-p_1)$ is just the momentum integral of eq. \ref{A3h}.
\enum

In fact, all hoop diagrams for any $\bar{\Upsilon}\,V^n\,\Upsilon$ vertices are finite because of the non-cancellation of `sufficient' momentum factors in the denominator.

\subsubsect{\texorpdfstring{$\bar{\Upsilon}\,\Upsilon\,\bar{\Upsilon}\,\Upsilon$}{Scalar 4-point Function}}
Such a vertex does not appear in the action and hence, the 1-hoop diagrams (figure \ref{S1H4P}) contributing to this vertex can not be divergent.
\bfig[h]\bc
\includegraphics[width=12cm,height=2.2cm]{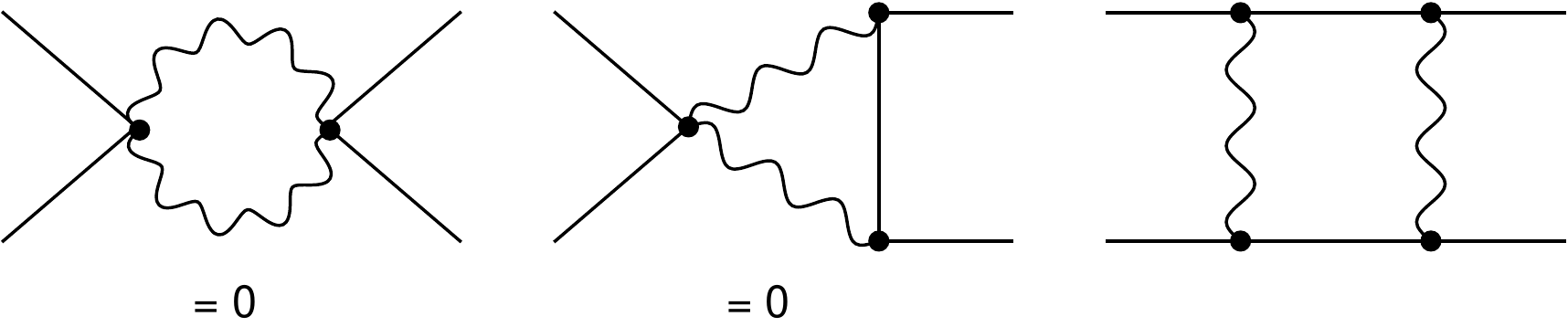}
\caption{$\bar{\Upsilon}\,\Upsilon\,\bar{\Upsilon}\,\Upsilon$ diagrams at 1-hoop.}
\label{S1H4P}
\ec\efig

Out of the three diagrams, two vanish due to $d-$algebra and the remaining box-diagram can be evaluated in a straightforward manner to give a finite result:
\be
\sim\int\frac{d^4k}{(2\pi)^4} \frac{1}{\frac{1}{2}k^2\frac{1}{2}(k+p_2)^2\frac{1}{2}(k+p_2+p_3)^2\frac{1}{2}(k-p_1)^2}\int d^8\theta\int \frac{dy_{1,2}}{y_1y_2}\frac{\bar{\Upsilon}_1(p_1){\Upsilon}_1(p_2)\bar{\Upsilon}_2(p_3){\Upsilon}_2(p_4)}{y_{12}\,y_{21}}\label{A4h}
\ee

\subsubsect{Vector Self-energy}
There are three classes of diagrams contributing to the self-energy corrections with different hyperfields (vector, ghosts or scalar) running inside the loop as shown in figure \ref{V1H2P}:
\bfig[h]\bc
\includegraphics[width=14cm,height=2.75cm]{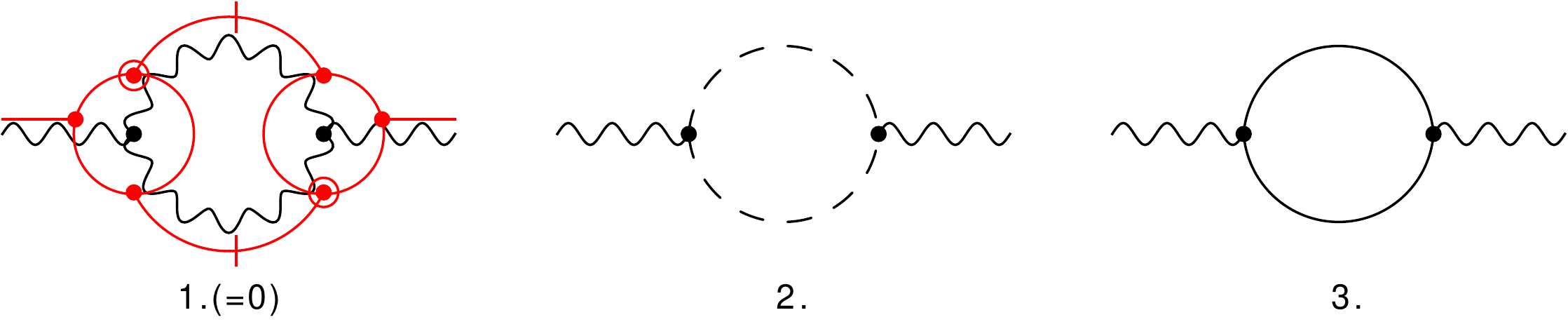}
\caption{Vector self-energy diagrams at 1-hoop.}
\label{V1H2P}
\ec\efig
\bnum
\item $V-$propagators: There are a couple of diagrams (not shown explicitly in fig. \ref{V1H2P}.1) which have at least one $V_1^2V_2$-type vertex and they vanish trivially due to the presence of expressions like $y^3\delta(y)$ or $y^4\delta(y)^2$. This is a generic feature of 1-hoop (at least) diagrams containing such vertices and these will not be considered here anymore.

The diagram which has both vertices of $V_1V_2V_3$-type (shown explicitly in fig. \ref{V1H2P}.1\footnote{The red (straight) lines over the wavy lines depict the `$y-$dependence' of the diagram following the rules given in Appendix \ref{App:yC}.}) also vanishes but in a different way. After doing the $d-$algebra and integrating the two $\delta(y)$'s, we are left with the following $y-$integrals:
\begin{align*}
{}&\int dy_{1,2,a,b}\frac{V_1V_2\,y_{ab}\,y_{ba}}{y_a\,y_b\,y_{1b}\,y_{a1}\,y_{2a}\,y_{b2}}\\
=&\int dy_{1,2}\frac{V_1V_2}{y_{12}y_{21}}\int dy_{a,b}\left(\frac{1}{y_{a1}}+\frac{1}{y_{2a}}\right)\left(\frac{1}{y_{1b}}+\frac{1}{y_{b2}}\right)\left(2-\frac{y_a}{y_b}-\frac{y_b}{y_a}\right)\\
=&\int dy_{1,2}\frac{V_1V_2}{y_{12}y_{21}}\left(2-\frac{y_1}{y_1}-\frac{y_2}{y_2}\right)=0.
\end{align*}

\item $(b,c)-$propagators: There are four diagrams with different combinations of ghost propagators and vertices. All of them combine to give (after relevant $d-$algebra)\footnote{$V\rightarrow gV$ in rest of the paper.}:
\begin{align}
{}&\,\A_2(p)\times 2\frac{1}{2}c_A\frac{g^2}{4}\int d^8\theta\int dy_{1,2}\frac{V_1V_2}{y_{12}\,y_{21}}\left(1+\frac{y_1}{y_2}+\frac{y_2}{y_1}+1\right)\nonumber \\
=&\,\A_2(p)\times\frac{c_A}{4}g^2\int d^8\theta\int dy_{1,2}\frac{V_1V_2}{y_{12}\,y_{21}}\left(\frac{-y_{12}\,y_{21}+4y_1y_2}{y_1y_2}\right)\nonumber \\
=&\,\A_2(p)\times c_A\,g^2\int d^8\theta\int dy_{1,2}\frac{V_1V_2}{y_{12}\,y_{21}}.\label{V1H2Pv}
\end{align}
The last line follows entirely from the second term in parentheses of the previous line. This is because the first term with no $y_{12}$'s in the integrand vanishes since $d^8\theta$ kills such projective integrands. Also, $\A_2(p)$ is the divergent integral and is evaluated using dimensional regularization to give:
\[\A_{2}(p)=\int\frac{d^Dk}{(2\pi)^D}\frac{1}{\frac{1}{2}k^2\frac{1}{2}(k+p)^2}=\frac{1}{4\pi^2}\left[\frac{1}{\epsilon}-\gamma_E-ln\left(\frac{p^2}{\mu^2}\right)\right]\,,\quad \frac{1}{\epsilon}=\frac{2}{4-D}\,\]

\item $\Upsilon-$propagators: The calculation for this single diagram is similar to that of the ghost which gives:
\be
-\A_2(p)\times c_R\,g^2\int d^8\theta\int dy_{1,2}\frac{V_1V_2}{y_{12}\,y_{21}}.
\label{V1H2Ps}
\ee
\enum

\subsubsect{\texorpdfstring{$V_1\,V_2\,V_3$}{Vector 3-point Function}}
Similar to the vector self-energy case, three classes of diagrams contribute in this case also as shown in figure \ref{V1H3P}. We give only the final results after doing the $d-$algebra and $y-$calculus in what follows.
\bfig[h]\bc
\includegraphics[width=14cm,height=3cm]{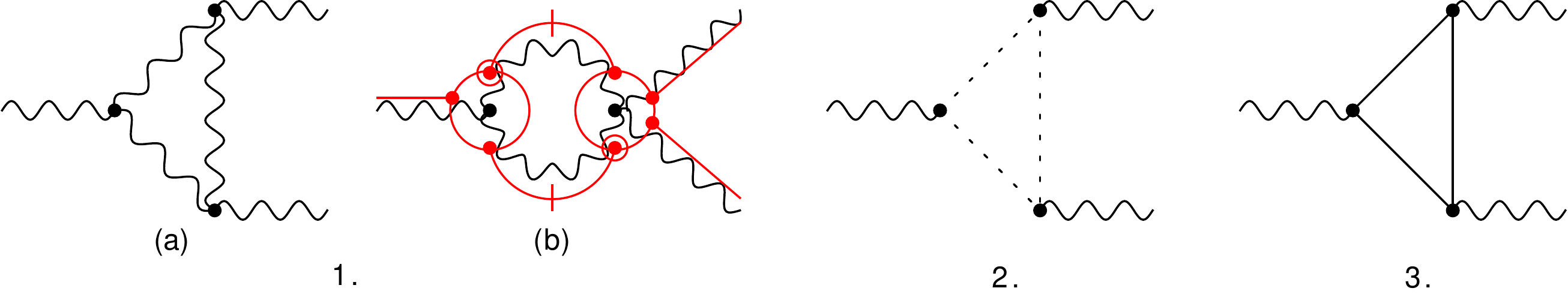}
\caption{$V_1\,V_2\,V_3$ diagrams at 1-hoop.}
\label{V1H3P}
\ec\efig
\bnum
\item $V-$propagators: There are two diagrams in this class and both are non-zero. We use the notation $\left\updownarrow\y{i}{j}\right\updownarrow$ to denote the sum of permutations of $\y{i}{j}-$factors over all possible values of $i$, e.g. $\left\updownarrow\y{1}{2}\right\updownarrow=\left(\y{1}{2}+\y{2}{3}+\y{3}{1}\right)$.
\bnum
\item $(VVV)^3$ vertices: The full (divergent \& finite) contribution of this diagram reads:
\begin{align}
-\frac{c_A}{2}g^3\int d^8\theta\int dy_{1,2,3}&\frac{V_1V_2V_3}{y_{12}\,y_{23}\,y_{31}}\Biggl[\A_2(p_3)\left(-3+\left\updownarrow\y{1}{2}\right\updownarrow\right)+\nonumber\\
{}&+p_2^2\,\A_3(p_2,-p_1)\left(1+\left\updownarrow\y{1}{2}-\frac{1}{3}\left(\y{31}{1}\right)^2\right\updownarrow\right)\Biggr]\label{V33}
\end{align}

\item $(VVVV)-(VVV)$ vertices: The (wavy line) diagram looks like contributing only to the $V_1V_2^2$ vertex term in the action but due to the 4$-$point vertex, this diagram also contributes to $V_1V_2V_3$ vertex as shown explicitly by the `$y$-dependence' in figure \ref{V1H3P}.1.(b). We will be giving the results for diagrams with all (external) $V$'s at distinct $y$'s only since the results for other diagrams follow from those of the self-energy case. This particular diagram gives a very simple contribution similar to the self-energy case:
\be
-\A_2(p_3)\times \frac{1}{2}\frac{c_A}{2}g^3\int d^8\theta\int dy_{1,2,3}\frac{V_1V_2V_3}{y_{12}\,y_{23}\,y_{31}}\left(3-\left\updownarrow\y{1}{2}\right\updownarrow\right)\label{V4V3}
\ee 
\enum

\item $(b,c)-$propagators: There are eight diagrams with different combinations of ghost propagators and vertices. All of them combine to give the following part containing the divergence:
\begin{align}
{}&\,\A_2(p_3)\times 2\frac{c_A}{2}\frac{g^3}{8}\int d^8\theta\int dy_{1,2,3}\frac{V_1V_2V_3}{y_{12}\,y_{23}\,y_{31}}\left(1+\left\updownarrow\y{1}{2}+\y{1}{3}\right\updownarrow +1\right)\nonumber\\
=&\,\A_2(p_3)\times \frac{c_A}{8}g^3\int d^8\theta\int dy_{1,2,3}\frac{V_1V_2V_3}{y_{12}\,y_{23}\,y_{31}}2\left(1+\left\updownarrow\y{1}{2}\right\updownarrow-\frac{y_{12}\,y_{23}\,y_{31}}{2\,y_1\,y_2\,y_3}\right)\label{bc8}
\end{align}
where the last term in the parentheses does not contribute as in the case of self-energy diagram but the last term does contribute in the remaining finite part given below:
\be
p_2^2\,\A_3(p_2,-p_1)\times \frac{c_A}{4}g^3\int d^8\theta\int dy_{1,2,3}\frac{V_1V_2V_3}{y_{12}\,y_{23}\,y_{31}}\frac{1}{3}\left\updownarrow\left(\y{12}{31}\right)^2\right\updownarrow\left(1+\left\updownarrow\y{1}{2}\right\updownarrow-\frac{y_{12}\,y_{23}\,y_{31}}{2\,y_1\,y_2\,y_3}\right)\label{bc8f}
\ee

\item $\Upsilon-$propagators: The calculation for this single diagram is again straightforward and gives as expected:
\be
-c_R\,g^3\int d^8\theta\int dy_{1,2,3}\frac{V_1V_2V_3}{y_{12}\,y_{23}\,y_{31}}\left[\A_2(p_3)+p_2^2\,\A_3(p_2,-p_1)\frac{1}{3}\left\updownarrow\left(\y{12}{31}\right)^2\right\updownarrow\right].
\label{V1H3Ps}
\ee
\enum

\subsubsect{\texorpdfstring{$V_1\,V_2\,V_3\,V_4$}{Vector 4-point Function}}
The calculations in this case are similar to the earlier ones except for an increase in the number of $y-$integrals to be evaluated. Before we proceed further, we make a group theoretical comment. None of the 4$-$point diagrams generate terms proportional to \[f_{ipq}f_{jqr}f_{krs}f_{lsp}V_1^i V_2^j V_3^k V_4^l\equiv G_{ijkl}V_1^i V_2^j V_3^k V_4^l\,,\] which do not appear in the SYM action\footnote{Recall from sub-section \ref{V} that the $V^4$ term appearing in the action is proportional to $f_{ijp}f_{klp}V_1^i V_2^j V_3^k V_4^l$.}. This was, however, not the case when similar calculations were done using N=1 supergraph rules in \cite{Storey} and a nonlinear (cubic) wavefunction renormalization (proportional to $GVVV$) was required to keep the effective action renormalizable as predicted in \cite{PS}.

We do not encounter this feature because of the `antisymmetry' of the $y_{ab}$ factors, which enforces the Jacobi identity leading (in this particular case) to this useful identity: \[G_{ijkl}-G_{ijlk}=\frac{c_A}{2}f_{ijp}f_{klp}\equiv\frac{c_A}{2}\line(1,0){10}\line(0,1){10}\line(1,0){5}\line(1,0){6}\line(0,1){10}\line(1,0){10}\,.\] Hence, all the 4$-$point diagrams end up producing the $V^4$ term present in the SYM action and the usual renormalization procedure is applicable. (One more reason is that \emph{`$gV$' is not renormalized} as explained later.) Now, we enumerate the complete results for the usual three classes of diagrams shown in figure \ref{V1H4P}:
\bfig[h]\bc
\includegraphics[width=15cm,height=6cm]{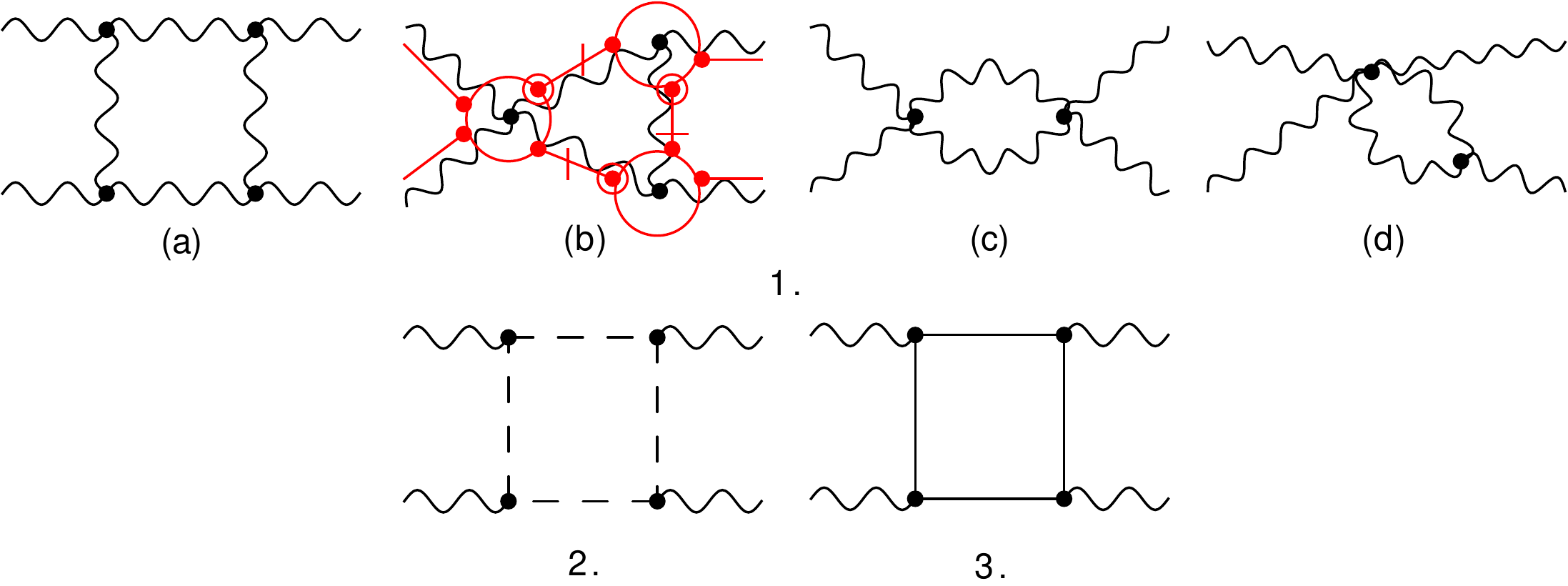}
\caption{$V_1\,V_2\,V_3\,V_4$ diagrams at 1-hoop.}
\label{V1H4P}
\ec\efig
\bnum
\item $V-$propagators: There are four non-zero diagrams in this class. After doing the relevant algebra and including the permutations of $\y{i}{j}-$factors, we get:
\bnum
\item $(VVV)^4$ vertices:
\begin{align}
&-\frac{c_A}{2}g^4\int d^8\theta\int dy_{1,2,3,4}\frac{V_1V_2V_3V_4}{y_{12}\,y_{23}\,y_{34}\,y_{41}}\Biggl[\left(\A_2(p_4)-p_2^2\A_3(-p_3,p_4)\right)\Biggl(\frac{1}{2}\left\updownarrow\y{1}{2}+\y{1}{4}\right\updownarrow-\nonumber\\
{}&-\left\updownarrow\y{1}{3}\right\updownarrow\Biggr)+\A_3(-p_3,p_4)\Biggl\{p_1^2\left(\frac{1}{4}\left\updownarrow 3\y{1}{2}-2\y{1}{3}-\frac{y_3^2}{y_1^2}-\frac{y_3^2}{y_1\,y_4}+\frac{y_3^3}{y_1^2\,y_4}\right\updownarrow\right)+\nonumber\\
{}&+p_2^2\left(-2+\frac{1}{4}\left\updownarrow 4\y{1}{2}-3\y{1}{3}+\frac{y_3\,y_4}{y_1^2}-\frac{y_4^2}{y_1^2}+\frac{y_4^2}{y_1\,y_2}\right\updownarrow\right)+\nonumber\\
{}&+2p_1\cdot p_2\left(2-\frac{1}{4}\left\updownarrow 4\y{1}{2}-\y{1}{3}+\frac{y_3^2}{y_1^2}-\frac{y_3^2}{y_1\,y_4}-\frac{y_3\,y_4}{y_1^2}\right\updownarrow\right)\Biggr\}+\nonumber\\
{}&+p_1^2(p_1+p_2)^2\A_4(p_2,p_2+p_3,-p_1)\left(-4+\frac{1}{4}\left\updownarrow 5\y{1}{2}+2\y{1}{3}-\frac{y_3^2}{y_1^2}-\frac{3y_3^2}{y_1\,y_4}+\frac{y_3^3}{y_1^2\,y_4}\right\updownarrow\right)\Biggr]\label{V34}
\end{align}

\item $(VVVV)-(VVV)^2$ vertices:
\begin{align}
-\frac{c_A}{2}g^4\int d^8\theta\int dy_{1,2,3,4}&\frac{V_1V_2V_3V_4}{y_{12}\,y_{23}\,y_{34}\,y_{41}}\Biggl[\A_2(p_4)\left(-8+2\left\updownarrow\y{1}{3}\right\updownarrow\right)+\nonumber\\
{}&+p_3^2\,\A_3(-p_3,p_4)\left\updownarrow\frac{y_{14}}{y_{12}}\frac{y_{24}^3}{y_2^2\,y_4}\right\updownarrow\Biggr]\label{V4V32}
\end{align}

\item $(VVVV)^2$ vertices:
\be
-\A_2(p_4)\times\frac{1}{2}\frac{c_A}{2}g^4\int d^8\theta\int dy_{1,2,3,4}\frac{V_1V_2V_3V_4}{y_{12}\,y_{23}\,y_{34}\,y_{41}}\left(4-\frac{1}{2}\left\updownarrow\y{1}{2}+\y{1}{4}\right\updownarrow\right)\label{V42}
\ee 

\item $(VVVVV)-(VVV)$ vertices:
\be
-\A_2(p_4)\times \frac{1}{2}c_A g^4\int d^8\theta\int dy_{1,2,3,4}\frac{V_1V_2V_3V_4}{y_{12}\,y_{23}\,y_{34}\,y_{41}}\left(4-\left\updownarrow\y{1}{3}\right\updownarrow\right)\label{V5V3}
\ee 
\enum

\item $(b,c)-$propagators: There are sixteen diagrams that combine to give (after dropping the term with a projective integrand):
\begin{align}
\frac{c_A}{16}g^4&\int d^8\theta\int dy_{1,2,3,4}\frac{V_1V_2V_3V_4}{y_{12}\,y_{23}\,y_{34}\,y_{41}}\Biggl[\left(\A_2(p_4)-p_2^2\A_3(-p_3,p_4)\right)\left(2\left\updownarrow\y{1}{2}+\y{1}{4}\right\updownarrow\right)+\nonumber\\
{}&+\Biggl\{\A_3(-p_3,p_4)\frac{1}{4}\left(p_1^2\left\updownarrow\left(\y{24}{41}\right)^2\right\updownarrow+p_2^2\left\updownarrow\left(\y{13}{41}\right)^2\right\updownarrow-2p_1\cdot p_2\left\updownarrow\y{23}{14}+\y{12}{41}\y{34}{41}\right\updownarrow\right)+\nonumber\\
{}&+p_1^2(p_1+p_2)^2\A_4(p_2,p_2+p_3,-p_1)\frac{1}{4}\left\updownarrow\left(\y{12}{41}\right)^2\right\updownarrow\Biggr\}\left(2\left\updownarrow\y{1}{2}+\y{1}{4}\right\updownarrow+\frac{y_{12}\,y_{23}\,y_{34}\,y_{41}}{y_1\,y_2\,y_3\,y_4}\right)\Biggr]\label{bc16}
\end{align}

\item $\Upsilon-$propagators: Without doing any new calculations, we can write the result, which is similar to eq. \ref{bc16}:
\begin{align}
-c_R\,g^4&\int d^8\theta\int dy_{1,2,3,4}\frac{V_1V_2V_3V_4}{y_{12}\,y_{23}\,y_{34}\,y_{41}}\Biggl[\left(\A_2(p_4)-p_2^2\A_3(-p_3,p_4)\right)+\nonumber\\
{}&+\A_3(-p_3,p_4)\frac{1}{4}\left(p_1^2\left\updownarrow\left(\y{24}{41}\right)^2\right\updownarrow+p_2^2\left\updownarrow\left(\y{13}{41}\right)^2\right\updownarrow-2p_1\cdot p_2\left\updownarrow\y{23}{14}+\y{12}{41}\y{34}{41}\right\updownarrow\right)+\nonumber\\
{}&+p_1^2(p_1+p_2)^2\A_4(p_2,p_2+p_3,-p_1)\frac{1}{4}\left\updownarrow\left(\y{12}{41}\right)^2\right\updownarrow\Biggr].\label{V1H4Ps}
\end{align}
\enum

\subsect{\texorpdfstring{1-hoop $\beta-$function}{1-hoop Beta-function}}
The divergences proportional to the terms in the vector hypermultiplet's action are absorbed via wavefunction ($V$) and coupling constant ($g$) renormalization following the usual well-known procedure.
\begin{align}
\mathrm{Z-factor\,for\,}V:&\quad V_R=\sqrt{Z_V} V\Rightarrow Z_V^{(1)}=Z_2^{(1)}\label{ZV}\\
\mathrm{Z-factor\,for\,}g:&\quad g_R=Z_g g\mu^\epsilon\Rightarrow Z_g^{(1)}=Z_3^{(1)}\left(Z_V^{(1)}\right)^\frac{-3}{2}\label{Zg}
\end{align}
where, the $Z_n^{(1)}$'s are the $Z-$factors for corresponding $n-$point vertex terms in the action, i.e. $\S(V^n_R)=Z_n\S(V^n)$. To figure these out, we combine the divergent term of $\A_2$ occurring in all $n-$point functions. The result is:
\begin{align}
2-\mathrm{point\,(\ref{V1H2Pv}\,\&\,\ref{V1H2Ps}):}&\quad\frac{(c_A-c_R)g^2}{4\pi^2\epsilon}\int d^8\theta\int dy_{1,2}\frac{V_1V_2}{y_{12}\,y_{21}};\nonumber\\
3-\mathrm{point\,(\ref{V33}-\ref{V1H3Ps}):}&\quad\frac{(c_A-c_R)g^3}{4\pi^2\epsilon}\int d^8\theta\int dy_{1,2,3}\frac{V_1V_2V_3}{y_{12}\,y_{23}\,y_{31}};\nonumber\\
4-\mathrm{point\,(\ref{V34}-\ref{V1H4Ps}):}&\quad\frac{(c_A-c_R)g^4}{4\pi^2\epsilon}\int d^8\theta\int dy_{1,2,3,4}\frac{V_1V_2V_3V_4}{y_{12}\,y_{23}\,y_{34}\,y_{41}}.\nonumber\\
\Rightarrow\mathrm{Z-factors\,for\,Vertices}:&\quad Z_2^{(1)}=Z_3^{(1)}=Z_4^{(1)}=1+\frac{(c_A-c_R)g^2}{4\pi^2\epsilon}\label{Z234}
\end{align}

Finally, plugging eq. \ref{Z234} in \ref{ZV} \& \ref{Zg}, we get:
\begin{align}
Z_V^{(1)}=1+\frac{(c_A-c_R)g^2}{4\pi^2\epsilon};\\
Z_g^{(1)}=1-\frac{(c_A-c_R)g^2}{8\pi^2\epsilon}.
\end{align}
Using the coupling constant renormalization factor, the $1-$loop $\beta-$function for N=2 SYM coupled to matter in representation $R$ is easily calculated:
\be
\beta^{(1)}_{N=2}=g^3\frac{\partial\left(\epsilon Z_g^{(1)}\right)}{\partial g^2}=-\frac{(c_A-c_R)g^3}{8\pi^2}\left(=-\frac{(2n-c_R)g^3}{8\pi^2}\quad\mathrm{for\,\,SU(n)}\right).
\ee
For N=4 SYM, it is trivial to see that the beta-function vanishes at 1-loop since the scalar hypermultiplet is in adjoint representation (so $c_R=c_A$) implying \[\beta^{(1)}_{N=4}=0.\]

\subsect{2-hoops Finiteness}
We recall that any $n-$point function involving external scalar hyperfields (including ghosts) can not be divergent  and hence the hyperfields $b$, $c$ \& $\Upsilon$ and other terms in actions \ref{sYV} \& \ref{sghf} are not renormalized. Thus, we need to calculate just the vector self-energy corrections to compute the $\beta-$function at two-hoops. We can read off the $Z_g-$factor from $g\,\bar{b}\,Vc$ (or $g\bar{\Upsilon}V\Upsilon$ in case of N=4 SYM) vertex at 2-hoops (which is true even in the case of 1-hoop as can be easily checked.):
\be
Z_g^{(2)}=\left(Z_V^{(2)}\right)^{-\frac{1}{2}}.
\label{Zgv}
\ee

In other words, $gV$ is not renormalized which means that the vector hyperfield $V$ can not have any non-linear renormalization since the coupling constant $g$ is always linearly renormalized. This is the same result as in the background field formalism as far as renormalization is concerned!
\bfig[h]\bc
\includegraphics[width=9.5cm,height=6.5cm]{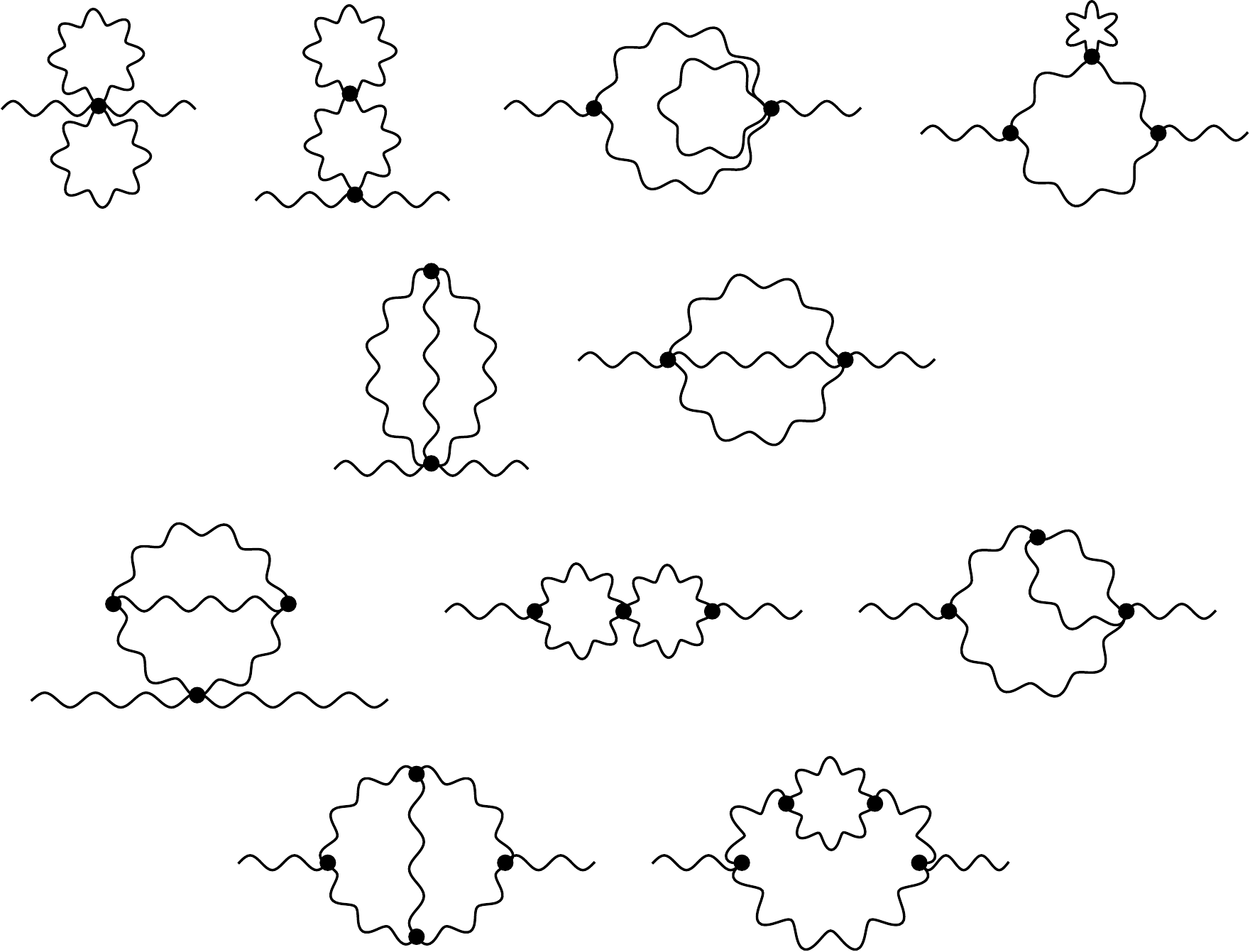}
\caption{Vector self-energy diagrams at 2-hoops with only $V-$propagators.}
\label{V2H2Pv}
\ec\efig
\bfig[h]\bc
\includegraphics[width=14cm,height=1.5cm]{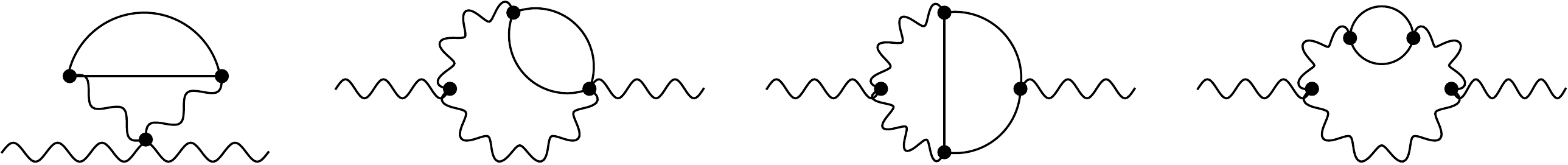}
\caption{Vector self-energy diagrams at 2-hoops including $b,\,c\,\&\,\Upsilon-$propagators.}
\label{V2H2Ps}
\ec\efig

There are a lot of diagrams to consider at 2-loops (at first sight) but their evaluation is not any more difficult than those at 1-loop. Firstly, we consider the 11 diagrams shown in figure \ref{V2H2Pv} which have only vector propagators. The first two rows in the figure show diagrams that vanish due to $d-$algebra (i.e. insufficient number of $d_{a\vartheta}^4$'s). The remaining 5 diagrams require some $y-$calculus and we find that none of their divergent terms survive and only the last two of them have finite terms. The vanishing of divergences for these two diagrams is shown in the appendix.

Secondly, there are a lot more diagrams having ghost \& scalar propagators but only 4 classes of such diagrams (figure \ref{V2H2Ps}) need to be examined in `detail'. The rest of such diagrams vanish either due to the $d-$algebra or emergence of $y\delta(y)$ (even $y^2\delta(y)$) factor (mainly in diagrams having only $\bar{b}\,Vc-$type vertices). Again, after doing some $y-$calculus we see that these four classes of diagrams also do not have any divergences. Thus, there are no divergent vector self-energy corrections at 2-hoops (i.e. $Z_V^{(2)}=1$) and hence for both N=2 SYM coupled to matter in any representation and N=4 SYM, \[\beta^{(2)}\sim \partial_{g^2}\left(Z_g^{(2)} \right)=0\,.\]

\newpage
\sect{Conclusion}
We investigated one \& two-loop(s) diagrams for N=2 massless vector \& scalar hypermultiplets directly in projective hyperspace for the cases of 2, 3 \& 4$-$point functions. We found that the effective action receives only 1-loop divergent corrections, which have the same form as the classical action. We also calculated all the 1-hoop finite pieces of the diagrams. Some of them are similar to the classical action modulo the momentum factors whereas others have extra $y-$factors, whose `non-linearity' prevents any simplification of the results. In spite of that, we derived the well-known result that the N=2 SYM coupled to matter is 2-loops finite. These calculations also enable us to show that N=4 SYM is finite at one \& two-loop(s).

Similar calculations can be done with the Harmonic hyperspace Feynman rules and the procedure is not that different or difficult. However, the repeated use of harmonic derivatives ($d_y\,\&\,d_{\bar{y}}$) to simplify the SU(2) invariant harmonics in order to do the SU(2) integrals is definitely cumbersome compared to `evaluating' some contour integrals on a complex plane as in the Projective hyperspace!

Our results like the linear wavefunction renormalization and the non-renormalization of `$gV$' have the same simplicity as expected from a background field formalism. So, it would be more interesting to construct a background field formalism for projective hyperspace that would definitely simplify these calculations and also (hopefully) give us insights into the origin of the Vector Hyperfield, V!

\section*{\bc\color{sect}Acknowledgements\ec}
This work is supported in part by National Science Foundation Grant No. PHY-0653342. DJ would like to thank Prerit Jaiswal for some useful discussions on Feynman diagrams.

\newpage
\appendix
\appsect{\texorpdfstring{$y-$Calculus}{y-Calculus}\label{App:yC}}
\paragraph{Important Identities\\}
These `simple' identities are useful for proving gauge invariance of the vector action, deriving component action of N=2 SYM, deriving vector propagator and evaluating $y-$integrals for Feynman diagrams:
\begin{align}
\frac{1}{y_{ij}y_{jk}}&=\frac{1}{y_{ik}}\left(\frac{1}{y_{ij}}+\frac{1}{y_{jk}}\right)\\
\frac{1}{y_{12}}+\frac{1}{y_{21}}&=-2\pi\i\,\delta(y_{12})\\
\frac{y_2}{y_{12}}+\frac{y_1}{y_{21}}-1&=2\pi\i\,\frac{y_1+y_2}{2}\delta(y_{12})
\end{align}

\paragraph{Sample Calculations\\}
Pictorial rules for setting up $y-$integrals are shown in figure \ref{yCal} and some examples of applying these rules are given in figures \ref{V2H2Ps0}, \ref{V1H3Pai} \& \ref{V2H2Pv2}.
\bfig[h]\bc
\includegraphics[width=6cm,height=3cm]{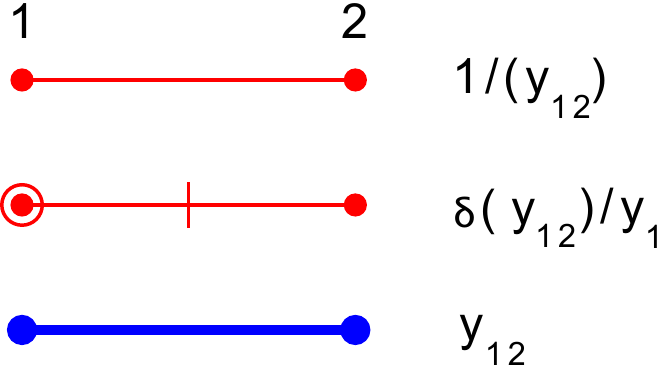}
\caption{Rules for setting up $y-$integrals.}
\label{yCal}
\ec\efig
\bfig[h]\bc
\includegraphics[width=14cm,height=2cm]{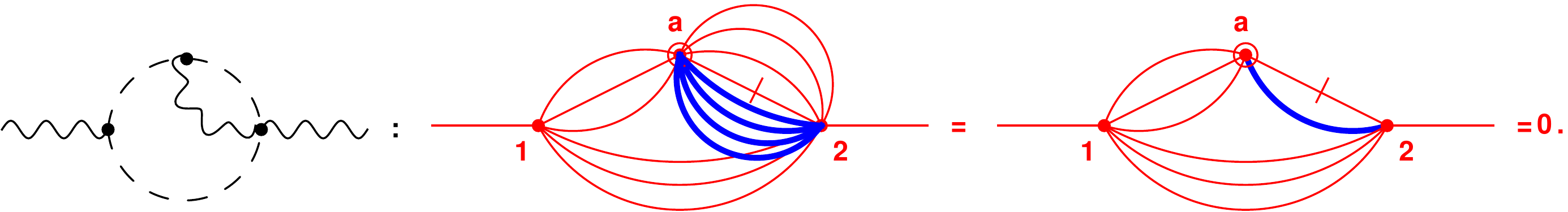}
\caption{Vanishing of a 2-hoops diagram with ghost propagators.}
\label{V2H2Ps0}
\ec\efig

In fig. \ref{V2H2Ps0}, the emergence of $y\delta(y)$ factor is shown when only three $y_{2a}$'s in the $y_{2a}^4$ factor (produced via $d-$algebra) are cancelled by $y_{2a}^3$ factor present in the ghost propagator.

Actual evaluation of `$q$' $y-$integrals (for the divergent pieces) is possible in `$\leq q$' steps as shown in the following sample calculations.
\bfig[h]\bc
\includegraphics[width=9cm,height=4cm]{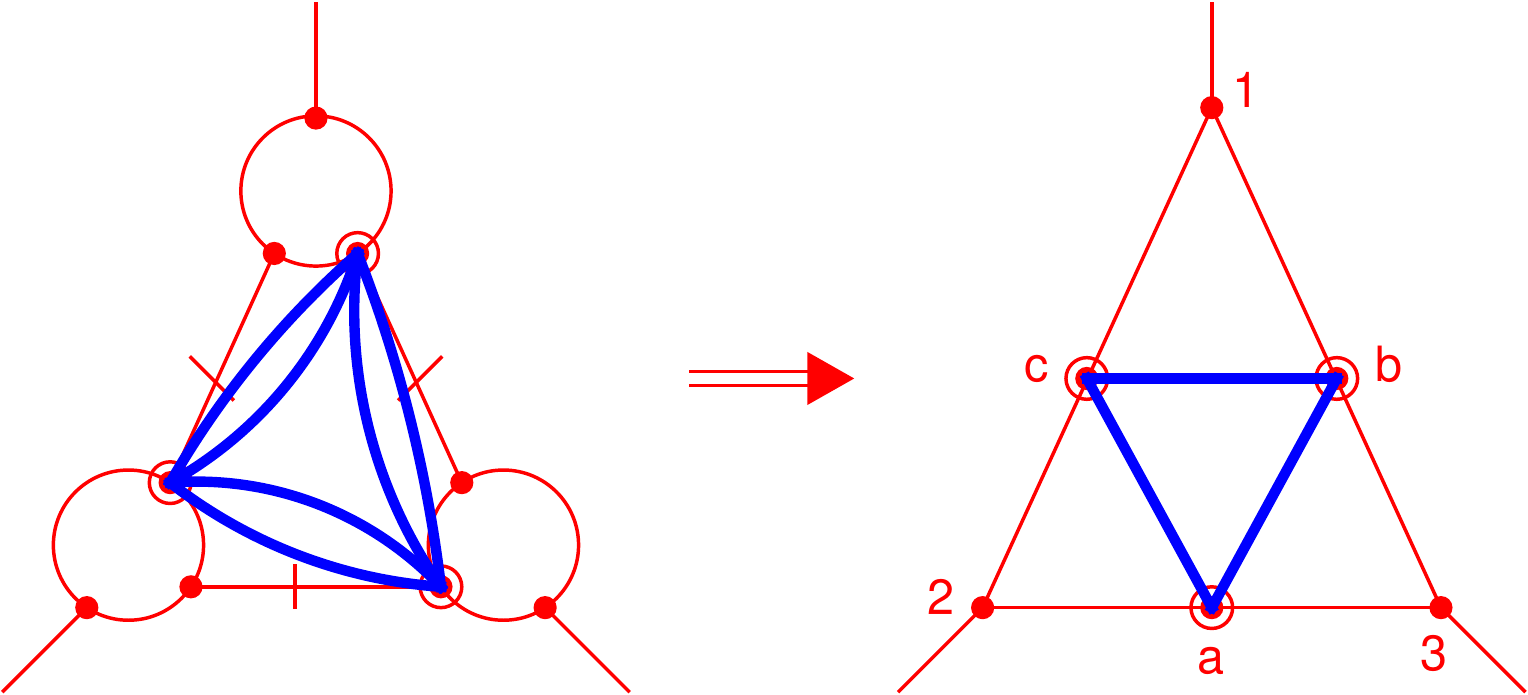}
\caption{Setting up $y-$integrals for diagram 1.(a) in fig. \ref{V1H3P}.}
\label{V1H3Pai}
\ec\efig
\begin{align*}
\mathrm{F.\ref{V1H3Pai}}\equiv&\int\frac{dy_{1,2,3,a,b,c}}{y_a\,y_b\,y_c}\frac{V_1V_2V_3\,y_{ba}\,y_{ac}\,y_{cb}}{y_{1c}\,y_{c2}\,y_{2a}\,y_{a3}\,y_{3b}\,y_{b1}}\\
=&\int dy_{1,2,3}\frac{V_1V_2V_3}{y_{12}\,y_{23}\,y_{31}}\oint dy_{a,b,c}\left(\frac{1}{y_{2a}}+\frac{1}{y_{a3}}\right) \left(\frac{1}{y_{3b}}+\frac{1}{y_{b1}}\right) \left(\frac{1}{y_{1c}}+\frac{1}{y_{c2}}\right) \times\\
{}&\qquad\qquad\times\left(1-\y{a}{b}\right) \left(1-\y{c}{a}\right) \left(1-\y{b}{c}\right)\\
=&\int dy_{1,2,3}\frac{V_1V_2V_3}{y_{12}\,y_{23}\,y_{31}}\oint dy_{a,b}\left(\frac{1}{y_{2a}}+\frac{1}{y_{a3}}\right) \left(\frac{1}{y_{3b}}+\frac{1}{y_{b1}}\right)\left(1-\y{a}{b}\right) \left(1-\y{2}{a}-\y{b}{1}+\y{b}{a}\right)\\
=&\int dy_{1,2,3}\frac{V_1V_2V_3}{y_{12}\,y_{23}\,y_{31}}\oint dy_{a}\left(\frac{1}{y_{2a}}+\frac{1}{y_{a3}}\right)\left(-\y{a}{3}-\y{2}{a}+\y{2}{3}+\y{a}{1}+\y{1}{a}-1\right)\\
=&\int dy_{1,2,3}\frac{V_1V_2V_3}{y_{12}y_{23}y_{31}}\left(-3+\left\updownarrow\y{1}{2}\right\updownarrow\right).
\end{align*}
\bfig[h]\bc
\includegraphics[width=12cm,height=8cm]{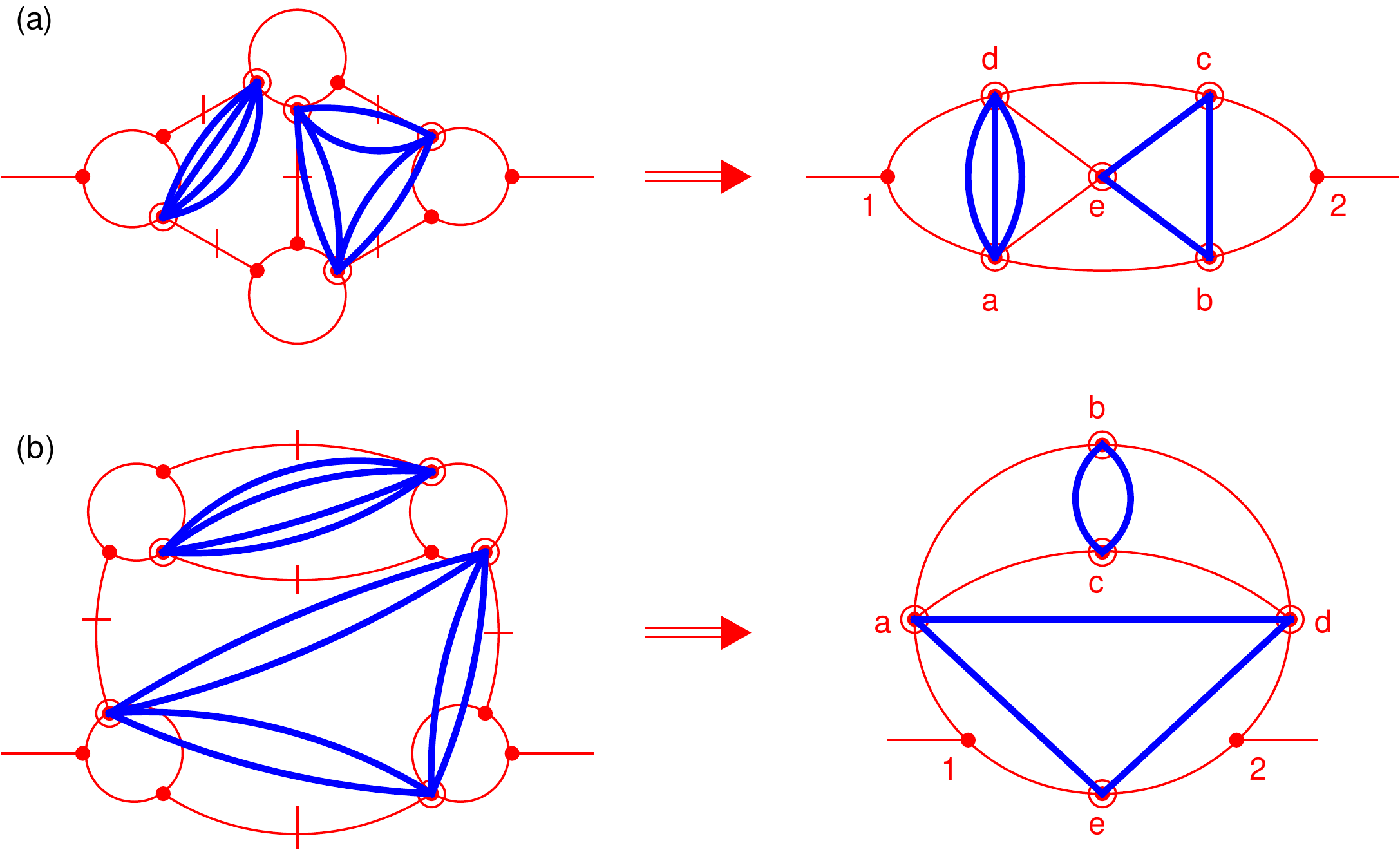}
\caption{Setting up $y-$integrals for the last two diagrams in fig. \ref{V2H2Pv}.}
\label{V2H2Pv2}
\ec\efig
\begin{align*}
\mathrm{F.\ref{V2H2Pv2}.(a)}\equiv&\int\frac{dy_{1,2,a,b,c,d,e}}{y_a\,y_b\,y_c\,y_d\,y_e}\frac{V_1V_2\,y_{ad}^3\,y_{bc}\,y_{ce}\,y_{eb}}{y_{1a}\,y_{ab}\,y_{b2}\,y_{2c}\,y_{cd}\,y_{de}\,y_{ea}\,y_{d1}}\\
=&\int\frac{dy_{1,2,a,b,c,d}}{y_a\,y_b\,y_d}\frac{V_1V_2\,y_{ad}^2\,y_{bc}}{y_{1a}\,y_{ab}\,y_{b2}\,y_{2c}\,y_{cd}\,y_{d1}}\left(1-\y{a}{c}-\y{b}{d}+\y{b}{c}\right)\\
=&\int\frac{dy_{1,2,a,b,d}}{y_a\,y_d}\frac{V_1V_2\,y_{ad}^2}{y_{1a}\,y_{ab}\,y_{b2}\,y_{d1}\,y_{2d}}\left(1-\y{a}{d}-\y{b}{d}+\y{b}{2}-\y{d}{b}+\y{a}{b}\right)\\
=&\int\frac{dy_{1,2,a,d}}{y_a\,y_d}\frac{V_1V_2\,y_{ad}^2}{y_{1a}\,y_{d1}\,y_{2d}\,y_{a2}}\left(3-\y{a}{d}-\y{2}{d}-\y{d}{a}\right)\\
=&\int dy_{1,2,a}\frac{V_1V_2}{y_{1a}\,y_{a2}\,y_{21}}\left(-6+4\y{a}{2}+5\y{1}{a}-\left(\y{a}{2}\right)^2-\y{2}{a}-\left(\y{1}{a}\right)^2\right)\\
=&\int dy_{1,2}\frac{V_1V_2}{y_{12}\,y_{21}}\frac{1}{2}\left(2-\left\updownarrow\y{1}{2}\right\updownarrow\right)\\
\Rightarrow&\frac{1}{2}\int d^8\theta\int dy_1dy_2\frac{V_1V_2}{y_1\,y_2}=0.
\end{align*}
\begin{align*}
\mathrm{F.\ref{V2H2Pv2}.(b)}\equiv&\int\frac{dy_{1,2,a,b,c,d,e}}{y_a\,y_b\,y_c\,y_d\,y_e}\frac{V_1V_2\,y_{bc}^2\,y_{ae}\,y_{ed}\,y_{da}}{y_{1a}\,y_{ba}\,y_{ac}\,y_{cd}\,y_{db}\,y_{d2}\,y_{2e}\,y_{e1}}\\
=&-\int\frac{dy_{1,2,a,b,d,e}}{y_a\,y_d\,y_e}\frac{V_1V_2\,y_{ae}\,y_{ed}}{y_{1a}\,y_{ba}\,y_{db}\,y_{d2}\,y_{2e}\,y_{e1}}\left(-2+\y{b}{a}-\y{d}{b}\right)\\
=&-\int\frac{dy_{1,2,a,d,e}}{y_a\,y_d\,y_e}\frac{V_1V_2\,y_{ae}\,y_{ed}}{y_{1a}\,y_{d2}\,y_{2e}\,y_{e1}}\left(-2+1+1\right)=0.
\end{align*}

\references{
\bibitem{ULMR}
U. Lindstr\"{o}m and M. Ro\v{c}ek, \cmp{115}{1988}{21};\\
U. Lindstr\"{o}m and M. Ro\v{c}ek, \cmp{128}{1990}{191}.
\bibitem{FGR}
F. Gonzalez-Rey, M. Ro\v{c}ek, S. Wiles, U. Lindstr\"{o}m and R. von Unge, \npb{516}{1998}{426} [\arXivid{hep-th/9710250}];\\
F. Gonzalez-Rey and R. von Unge, \npb{516}{1998}{449} [\arXivid{hep-th/9711135}];\\
F. Gonzalez-Rey, [\arXivid{hep-th/9712128}].
\bibitem{FGRc}
F. Gonzalez-Rey and M. Ro\v{c}ek, \pl{B}{434}{1998}{303} [\arXivid{hep-th/9804010}].
\bibitem{GIKOS}
A. Galperin, E. Ivanov, S. Kalitzin, V. Ogievetsky and E. Sokatchev, \cqg{1}{1984}{469};\\
A. Galperin, E. Ivanov, V. Ogievetsky and E. Sokatchev, {\it JETP\ Lett.\ } {\bf 40} (1984) 912 [{\it Pisma\ Zh.\ Eksp.\ Teor.\ Fiz.\ } {\bf 40} (1984) 155];\\
B. M. Zupnik, {\it Theor.\ Math.\ Phys.\ } {\bf 69} (1986) 1101 [{\it Teor.\ Mat.\ Fiz.\ } {\bf 69} (1986) 207];\\
A.S. Galperin, E.A. Ivanov, V.I. Ogievetsky, and E.S. Sokatchev, {\it Harmonic superspace} (Cambridge Univ. Press, 2001).
\bibitem{GIKOSc}
E. Ivanov, A. Galperin, V. Ogievetsky and E. Sokatchev, \cqg{2}{1985}{601};\\
E. Ivanov, A. Galperin, V. Ogievetsky and E. Sokatchev, \cqg{2}{1985}{617}.
\bibitem{BB-ASef}
I. Buchbinder, E. Buchbinder, E. Ivanov, S. Kuzenko and B. Ovrut, \pl{B}{412}{1997}{309} [\arXivid{hep-th/9703147}]. 
\bibitem{BB-Bg}
I. Buchbinder, E. Buchbinder, S. Kuzenko and B. Ovrut, \pl{B}{417}{1998}{61} [\arXivid{hep-th/9704214}];\\
I. Buchbinder, S. Kuzenko and B. Ovrut, \pl{B}{433}{1998}{335} [\arXivid{hep-th/9710142}].
\bibitem{BB-Bgc}
I. Buchbinder, E. Ivanov and A. Petrov, \npb{653}{2003}{64} [\arXivid{hep-th/0210241}].
\bibitem{DPHfH} 
D. Jain and W. Siegel, \pr{D}{80}{2009}{045024} [\arXivid{0903.3588}].
\bibitem{WS-AdS}
W. Siegel, 2010, \arXivid{1005.2317}.
\bibitem{S3B0}
M. Grisaru, M. Ro\v{c}ek and W. Siegel, \prl{45}{1980}{1063};\\
M. Grisaru, M. Ro\v{c}ek and W. Siegel, \npb{183}{1981}{141}.
\bibitem{CZ}
W. E. Caswell and D. Zanon, \pl{B}{100}{1981}{152};\\
W. E. Caswell and D. Zanon, \npb{182}{1981}{125}.
\bibitem{finite}
M. T. Grisaru and W. Siegel, \npb{201}{1982}{292};\\
P. S. Howe, K. S. Stelle and P. K. Townsend, \npb{236}{1984}{125}.
\bibitem{PS}
O. Piguet and K. Sibold, \npb{197}{1982}{257 \& 272}.
\bibitem{Storey}
J. W. Juer and D. Storey, \pl{B}{119}{1982}{125};\\
J. W. Juer and D. Storey, \npb{216}{1983}{185}.}
\end{document}